\documentclass[]{./cls_bib/pasj02} 
\usepackage[switch,mathlines]{lineno} 
\usepackage[round,authoryear]{natbib} 
\usepackage{color, comment, ascmac}
\usepackage[colorlinks=true,linktocpage=true,citecolor=blue,linkcolor=black,urlcolor=blue]{hyperref}



\usepackage{booktabs}
\usepackage{threeparttable}
\usepackage{array}
\newcolumntype{C}[1]{>{\centering\arraybackslash}p{#1}}

\jyear{2025}
\Received{2025 November 18}
\Accepted{2026 February 22}

\graphicspath{{./}} 

\begin{document} 

\title{ Resonant scattering at the center of the galaxy cluster PKS 0745-191 with XRISM }

\author{
Keita \textsc{Tanaka},\altaffilmark{1,2}\altemailmark\orcid{0009-0001-7796-9536} \email{keita.tanaka@ac.jaxa.jp}
Megan \textsc{Eckart},\altaffilmark{3}\email{}
Kotaro \textsc{Fukushima},\altaffilmark{1}
Liyi \textsc{Gu},\altaffilmark{4}
Kyoko \textsc{Matsushita},\altaffilmark{5}
Brian \textsc{McNamara},\altaffilmark{6}
Fran\c{c}ois \textsc{Mernier},\altaffilmark{7,8,9,10}
Ikuyuki \textsc{Mitsuishi},\altaffilmark{11}
Frederick \textsc{S. Porter},\altaffilmark{12}
Kosuke \textsc{Sato},\altaffilmark{13}
Makoto \textsc{Sawada},\altaffilmark{14}
Kazunori \textsc{Suda},\altaffilmark{5}
Irina \textsc{Zhuravleva},\altaffilmark{15}
and 
Noriko \textsc{Y. Yamasaki},\altaffilmark{1,2}\altemailmark\orcid{0000-0003-4885-5537}\email{yamasaki.noriko@jaxa.jp}
}

\altaffiltext{1}{Institute of Space and Astronautical Science (ISAS), Japan Aerospace Exploration Agency (JAXA), 3-1-1 Yoshinodai, Chuo-ku, Sagamihara, Kanagawa 252-5210, Japan}
\altaffiltext{2}{Department of Astronomy, Graduate School of Science, The University of Tokyo, 7-3-1 Hongo, Bunkyo-ku, Tokyo 113-0033, Japan}
\altaffiltext{3}{Lawrence Livermore National Laboratory, 7000 East Avenue, Livermore, CA, 94550, USA}
\altaffiltext{4}{SRON Netherlands Institute for Space Research, Leiden, The Netherlands}
\altaffiltext{5}{Faculty of Physics, Tokyo University of Science, Tokyo 162-8601, Japan}
\altaffiltext{6}{Department of Physics \& Astronomy, Waterloo Centre for Astrophysics, University of Waterloo, Ontario N2L 3G1, Canada}
\altaffiltext{7}{NASA / Goddard Space Flight Center, Greenbelt, MD 20771, USA}
\altaffiltext{8}{IRAP, CNRS, Université de Toulouse, CNES, UT3-UPS, Toulouse, France}
\altaffiltext{9}{Department of Astronomy, University of Maryland, 4296 Stadium Dr., College Park, MD 20742-2421, USA}
\altaffiltext{10}{Center for Research and Exploration in Space Science and Technology, NASA/GSFC (CRESST II), Greenbelt, MD 20771, USA}
\altaffiltext{11}{Department of Physics, Nagoya University, Aichi 464-8602, Japan}
\altaffiltext{12}{NASA / Goddard Space Flight Center, Greenbelt, MD 20771, USA}
\altaffiltext{13}{Department of Astrophysics and Atmospheric Sciences, Kyoto Sangyo University, Kyoto 603-8555, Japan}
\altaffiltext{14}{Department of Physics, Rikkyo University, Tokyo 171-8501, Japan}
\altaffiltext{15}{Department of Astronomy and Astrophysics, The University of Chicago, Chicago, IL 60637, USA}




\maketitle

\begin{abstract}
We report an evidence of the resonant scattering effect at the center of the galaxy cluster PKS~0745$-$191 with XRISM.
We analyzed XRISM/Resolve commissioning-phase observations of the distant cluster PKS~0745$-$191 ($z = 0.103$) with a 54~ks exposure. The gain drift was corrected using the onboard modulated X-ray source (MXS), and spectra were extracted from the all pixels that were well illuminated by MXS, the core region (four central pixels, $\sim$100~kpc), and the surrounding region.
A single-temperature collisional ionization equilibrium (CIE) model fits the full field of fiew spectrum with $kT \approx 6$~keV and a turbulent velocity of $\approx$120~km~s$^{-1}$. From the core ($r < 50$~kpc) spectrum, we detected a $\approx$22\% suppression of the Fe\,\emissiontype{XXV} He$\alpha$ resonance (w) line relative to the CIE prediction.
We performed a Monte Carlo simulations to calculate the resonant scattering (RS)  effect using radial profiles from Chandra data. The RS-inferred turbulence agrees with that by Resolve line broadening, demonstrating that RS provides an independent and consistent constraint on ICM turbulence. These results highlight XRISM/Resolve’s potential for turbulence studies in galaxy clusters.
\end{abstract}


\section{Introduction}
The dynamical structure of clusters of galaxies is a direct consequence of the cosmological structure formation. Their mass and number density are sensitive to the primordial density fluctuation ($\sigma_{8}, n_{s}$), evolution of the scale factor ($a(t)$ and w) \citep[e.g.,][]{allen2011,mantz2014}.
These quantities can be measured by the hydrostatic mass estimation using the temperature and density profile of the intracluster medium (ICM) observed mainly in X-ray bands. The physical conditions of the ICM can be affected and biased from the ideal equilibrium state by several properties, such as viscosity, thermal conduction, and energy transport in the plasma \citep[e.g.,][]{parrish2012}. The core structure of the gravitational potential reflects the nature of the dark matter \citep[e.g.,][]{navarro1997}, in addition to the astrophysical effects of member galaxies and active galactic nucleus (AGN) \citep[e.g.,][]{peirani2017}. 
The turbulent velocity and bulk motion at the cluster core are crucial for understanding the kinetics at the cluster's center. The extreme accuracy of the line energy determination with Resolve \citep{ishisaki2025} on XRISM \citep{tashiro2025} is a new tool for studying the velocity field in the ICM \citep[e.g.,][]{xrismcollaboration2024}. Another unique tool is the observation of the resonant scattering (RS) of emission lines. \par
ICM is generally treated as optically thin, but in regions of high ion density, such as cluster cores, and for emission lines with high oscillator strength, such as resonance lines, the optical depth can become $\tau \geq 1$ \citep[e.g.,][]{gilfanov1987,churazov2010}. This effect manifests as a suppression of the resonance line flux, which can be observed in X-ray spectra.
Usually, we obtain the emission measure of the ICM from the projected flux and calculate the density, assuming that the depth of the line of sight is the same as the apparent size. 
The flux suppression due to RS is sensitive to ion column density and turbulent velocity among other parameters, allowing the turbulent velocity to be estimated independently of the line broadening measurement. RS is also sensitive to the anisotropy and scale of motions of the velocity dispersion \citep{zhuravleva2011}, providing a unique probe of the directional structure of turbulence in cluster cores. 

Observations of cold galaxy groups and galaxies using the XMM-Newton have confirmed the characteristic decrease in emission line intensity seen in RS. Comparing a model assuming turbulence with observational results indicates that the turbulent velocity in galaxies and galaxy groups is of the order of approximately 100 km$\rm {s}^{-1}$ when assuming uniform isotropy \citep[e.g.,][]{werner2009,deplaa2012,ogorzalek2017}.
Hitomi detected evidence of the RS effect in the Perseus cluster with a velocity dispersion of $\simeq 80-240 \ \rm km \ s^{-1}$ \citep{hitomicollaboration2018a}.
As the recent XRISM observations suggest subsonic ($ v<200 \ \rm km \ s^{-1}$) velocity dispersion and bulk motions in many clusters of galaxies, such as Centaurus clusters \citep{xrismcollaboration2025}.
Since RS is sensitive to turbulent velocity and to bulk motions on small scales in the ICM \citep{zhuravleva2011}, it serves as an effective probe for investigating the internal velocity structure.

As PKS~0745$-$191 has a high central density and high temperature, the RS may be observable with XRISM Resolve.
PKS~0745$-$191 is known as the brightest galaxy cluster at $z=0.1030$ with a 2--10 keV X-ray luminosity of $1.6\times10^{45}$ erg s$^{-1}$
due to a strong peak of the surface brightness at the center and has been studied by previous X-ray observatories 
such as Einstein \citep{fabian1985}, Chandra \citep{sanders2014}, Suzaku \citep{walker2012}, etc. 
The redshift of the central galaxy is measured $z=0.1028$ in optical band \citep{hunstead1978} and $z=0.1024$ with stellar continuum fitting \citep{gingras2024}.
The cluster exhibits clear signatures of AGN feedback, including X-ray cavities with sufficient mechanical energy to offset cooling \citep{sanders2014}, and disturbed nebular gas with complex kinematics aligned with the radio axis \citep{gingras2024}.

In this work, we present the first X-ray observation of PKS~0745$-$191 with Resolve onboard XRISM.
We describe our observations and data reduction in the next section, followed by the spectral analysis in Section \ref{sec:spec}. RS simulation is also presented in Section \ref{sec:RS_Sim}. We discuss the observation and simulation results in Section \ref{sec:discussion} and conclude in Section \ref{sec:conclusion}.
We assume a standard $\Lambda$-CDM cosmology with $H_{0}=$ 70 km s$^{-1}$ Mpc$^{-1}$, $q_{0}=0$, $\Omega_{\Lambda}=0.73$.

\section{Observation and Data Reduction}
\label{sec:obs}
\begin{figure}[t]
  \centering
  \includegraphics[width=0.48\textwidth]{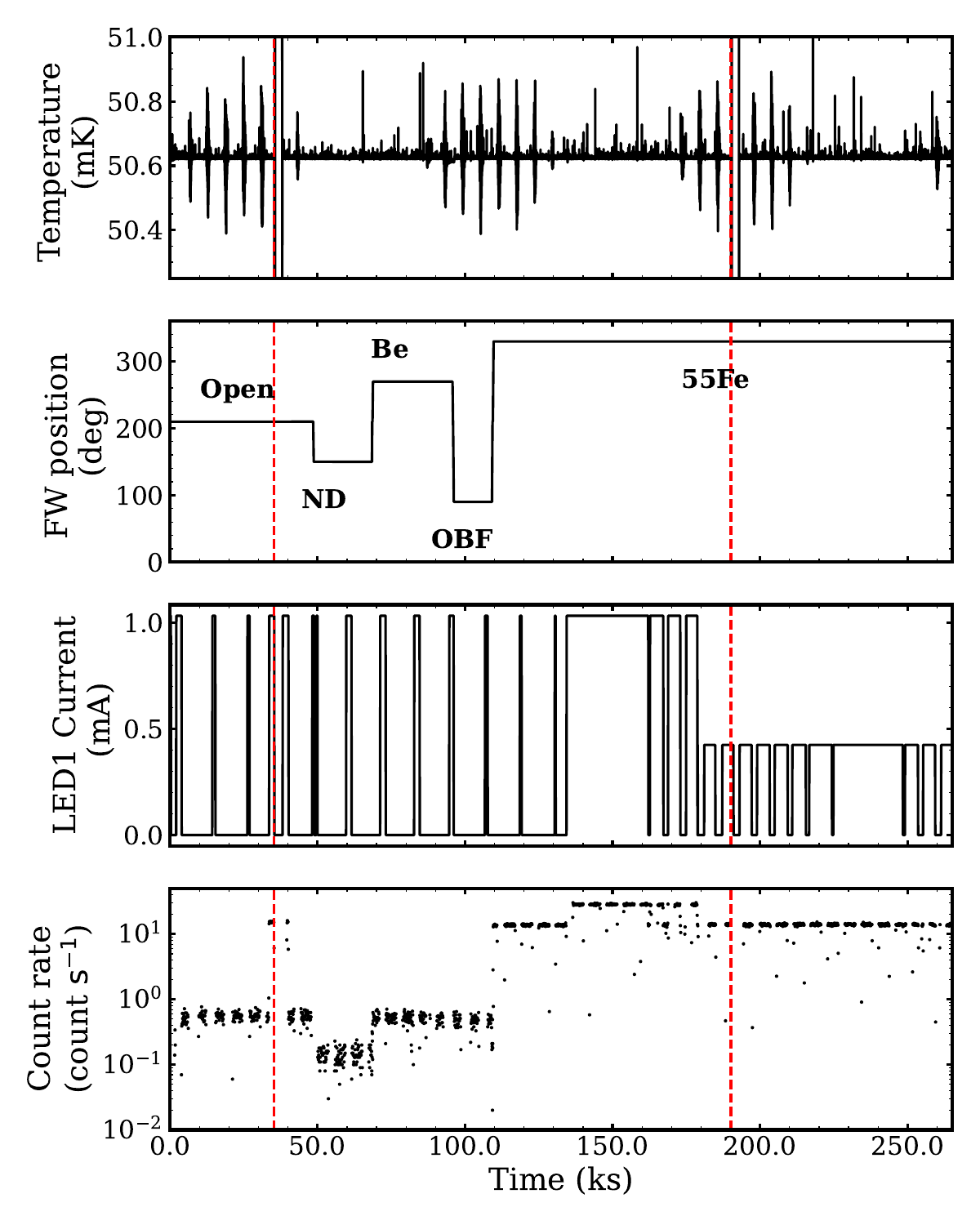}
  \caption{
  House keeping data during the PKS~0745$-$191 observation as a function of time. The origin of the time is 2023-11-08 10:21:04 UTC.
  From top to bottom: temperature of the cold stage of the adiabatic demagnetisation refrigerator (ADR), FW position, LED1 current driving the MXS, and 2--8~keV count rate of the Resolve array.  
  The red vertical lines mark the start time of ADR recycling.  
  The MXS LED current changed during the observation interval where $^{55}$Fe filter position was used.
  {Alt text: Four line graphs showing housekeeping data during the observation. Each panel shows time variations of ADR temperature, FW position, LED current, and count rate.}
  }
  \label{fig:FW-modes}
\end{figure}

XRISM observed the center of PKS~0745$-$191 during its commissioning phase in 8--11 November 2023 (OBSID 000112000).  
The observation was dedicated to commissioning the filter wheel (FW) and the modulated X-ray source (MXS).  
Resolve was operated FW with filter positions named OPEN (no filter), ND (neutral-density filter), Be (beryllium filter), OBF (optical-blocking filter) and $^{55}$Fe \citep{shipman2024} (Figure \ref{fig:FW-modes}) and, the MXS was intermittently irradiated with two parameter sets \citep{sawada2025} (Figure \ref{fig:FW-modes}).  
House-keeping data confirmed nominal FW and MXS behaviour.  
During this observation, the averaged heliocentric velocity is $22.685\pm 0.019\ \rm km \rm \ s^{-1}$ $(\Delta z = 7.57\times 10^{-5})$. 

\begin{figure}[t]
  \centering
  \includegraphics[width=0.48\textwidth]{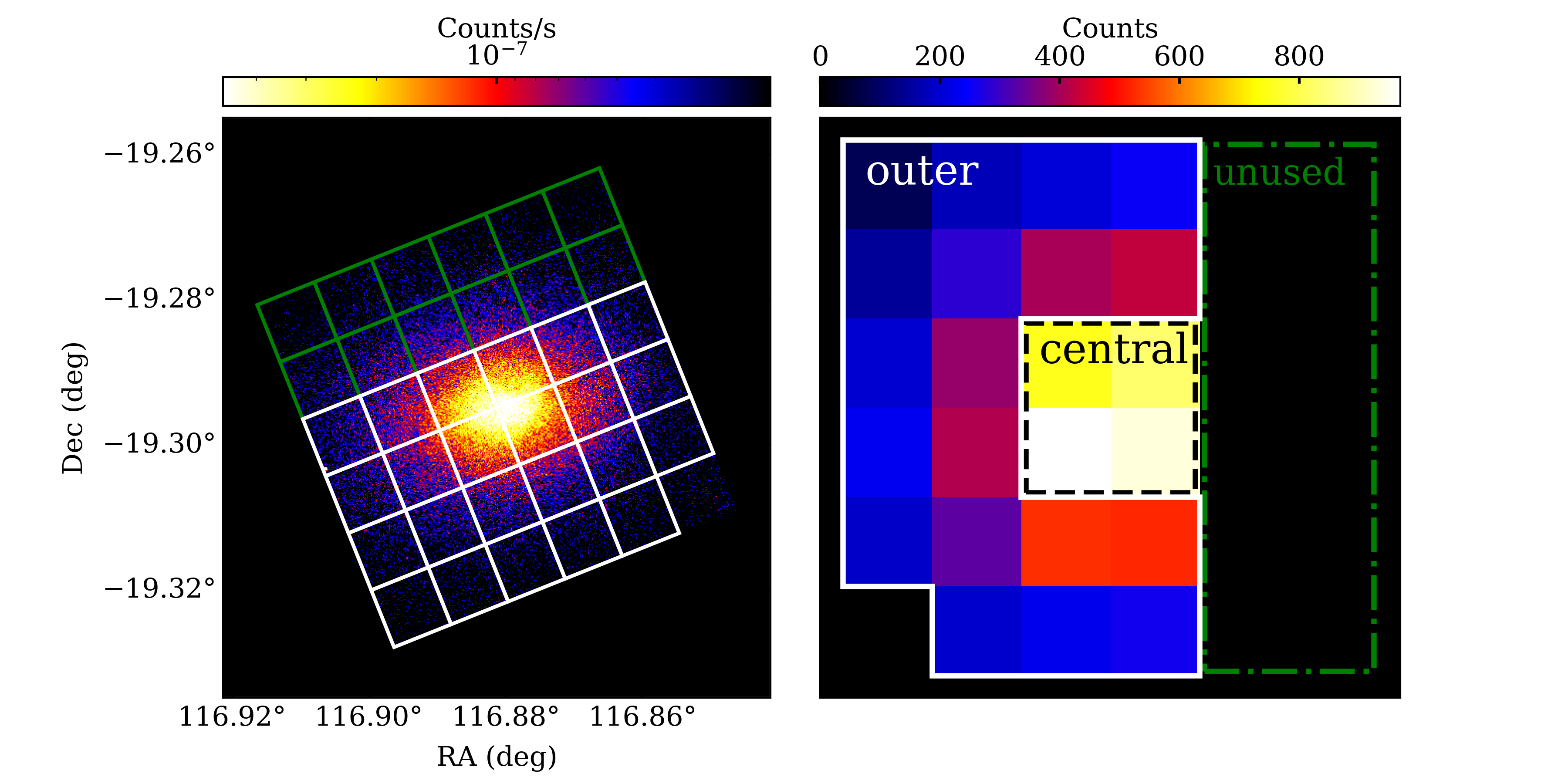}
  \caption{(Left): 2--8~keV  Chandra exposure-corrected image of PKS~0745$-$191 overlaid with the 23 Resolve pixels used in the analysis. The remaining 12 pixels are not used due to low MXS counts to correct the gain drift (green line). See the text for the details. (Right): Resolve 2--8~keV counts map. We divided the 23 Resolve pixels into central (central four pixels) and outer (other 19 pixels) regions.
  {Alt text: Two-panel color images of the Chandra and Resolve observations. The left panel is in sky coordinates (right ascension and declination), and the right panel is in detector coordinates.}} 
  \label{fig:image}
\end{figure}

During nominal science operations, the Resolve time dependent energy scale is reconstructed using periodic gain fiducials utilizing $^{55}$Fe radioactive sources on the Resolve FW (Porter et al., 2025). However, this observation was performed before the nominal observation strategy was implemented. For this observation, only the on-board MXS was available, giving us periodic gain fiducials using Cr K$\alpha$ at 5.4 keV. The energy scale reconstruction strategy for this observation is detailed in Appendix 1. In addition, the MXS shows a strong pixel-to-pixel gradient due to obstruction from the closed instrument gate valve structure. Thus, two full detector rows (see Figure 2) must be excluded from the analysis presented in this paper due to insufficient illumination from the MXS to reconstruct the detector gain and recover the energy scale. The estimated uncertainty in the energy scale in the band 5.4--8.0~keV is about 0.5 eV, varying slightly between the different FW positions used in the observation, as detailed in Appendix 1. The spectral resolution of the instrument for this observation is well represented by the instrument response function and is about 4.5 eV at 6 keV.

Our spectral analysis uses only the OPEN, ND, Be, and OBF data.
The FW $^{55}$Fe intervals were not used due to interference from the calibration source.
Data were processed with pipeline software Build 8, and the standard screening criteria (presented in XRISM ABC Guide\footnote{\parbox{0.9\linewidth}{\url{https://heasarc.gsfc.nasa.gov/docs/xrism/analysis/abc_guide/XRISM_Data_Analysis.html}.}}) were used.
The data to be analyzed does not include the time when the MXS LED is on.
The exposure time by Resolve with the OPEN, ND, OBF, and Be filters is 54 ks including the MXS irradiation periods. 
The Resolve calibration data were applied with CALDB~v8 and the file \texttt{xa\_rmfparam\_20190101v006.fits}, was used to generate the redistribution matrix files (RMFs).
The ancillary response files (ARFs) were generated using the \texttt{xaarfgen} task, assuming the surface brightness distribution derived from a Chandra image in the 2--8~keV energy band, with a field of view of $r=4'$ and a pixel size of $0.5"$.

Figure \ref{fig:image} shows the exposure-corrected image in \textit{Chandra} and the count map in Resolve FW open data.
One pixel ($0.5^\prime \times 0.5^\prime$) of Resolve corresponds to about $\sim 50$ kpc of PKS~0745$-$191, covering about $\sim 300$ kpc of the entire FoV.

\section{Spectral Analysis}
\label{sec:spec}
\subsection{Models}
We apply optimal binning \citep{kaastra2016} to Resolve spectra and fit them using \textsc{Xspec} v12.14.1 \citep{arnaud1996} with Cash statistics \citep{cash1979}.
The ICM model was calculated with \textsc{AtomDB} v3.0.9 \citep{foster2012,foster2020}.  
We modeled the ICM with \texttt{bvapec}, which attributes line broadening exclusively to velocity dispersion: the thermal broadening is fixed by the electron temperature, while any additional (non-thermal) broadening is described by an independent velocity-dispersion parameter.
The solar abundance was used \texttt{lpgs} in \textsc{Xspec} \citep{lodders2009} and the iron and nickel abundances were free to vary and the other elements were fixed to 0.4 $Z_{\odot}$.
The X-ray photoelectric absorption was modeled with a Galactic hydrogen column density of $N_H$ of $0.409\times 10^{-22}$ $\text{cm}^{-2}$ \citep{hi4picollaboration:2016} and using \texttt{TBabs} model \citep{wilms2000}. 
Non X-ray background (NXB) emission employed a phenomenological model \citep{xrismcollaboration2024} with variable normalization (see Appendix 2), to which diagonal responses were applied.

Because data of each FW position has different redshift due to the gain uncertainty $\sim 0.6\ \rm eV$ (Appendix 1), we simultaneously fitted the spectra of all filters, allowing the redshift to vary to compensate for the gain uncertainty.

\begin{figure}[t]
  \centering
  \begin{minipage}[b]{\linewidth}
    \centering
    \includegraphics[width=1.0\textwidth]{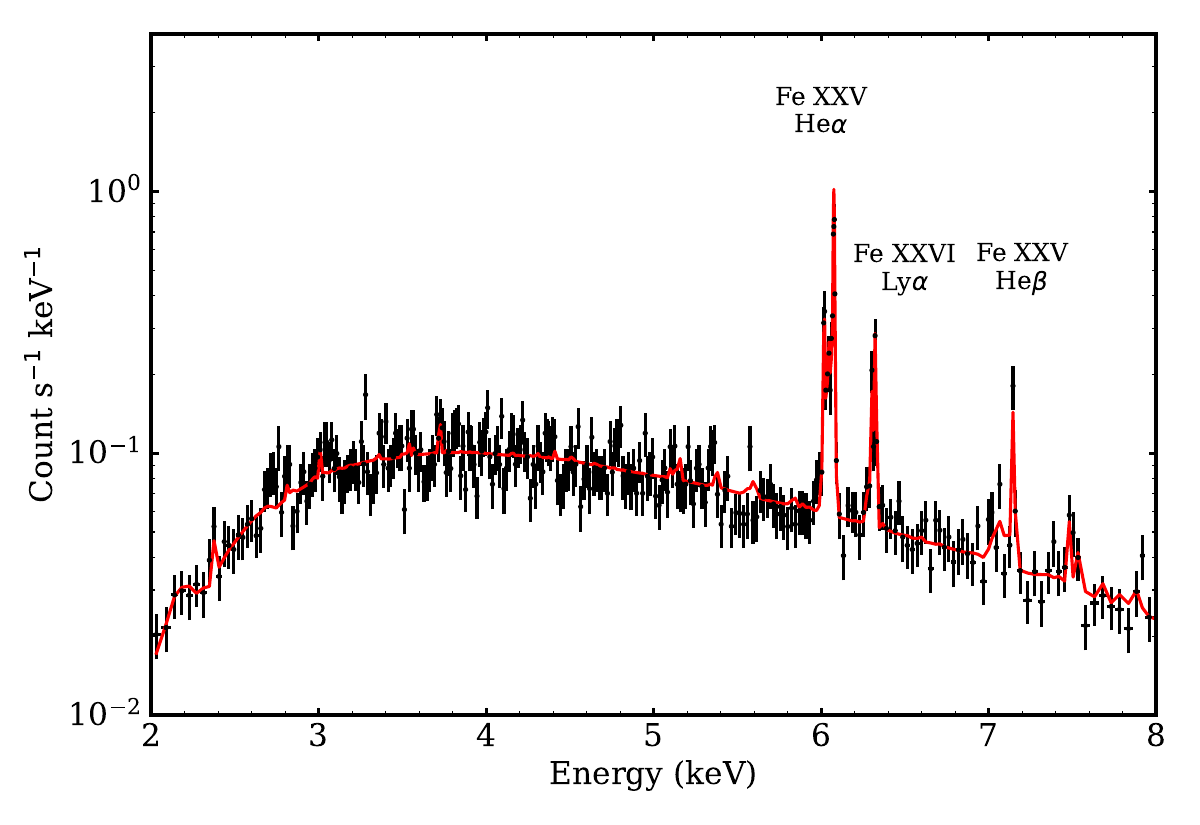}\\[1ex]
    \includegraphics[width=1.0\textwidth]{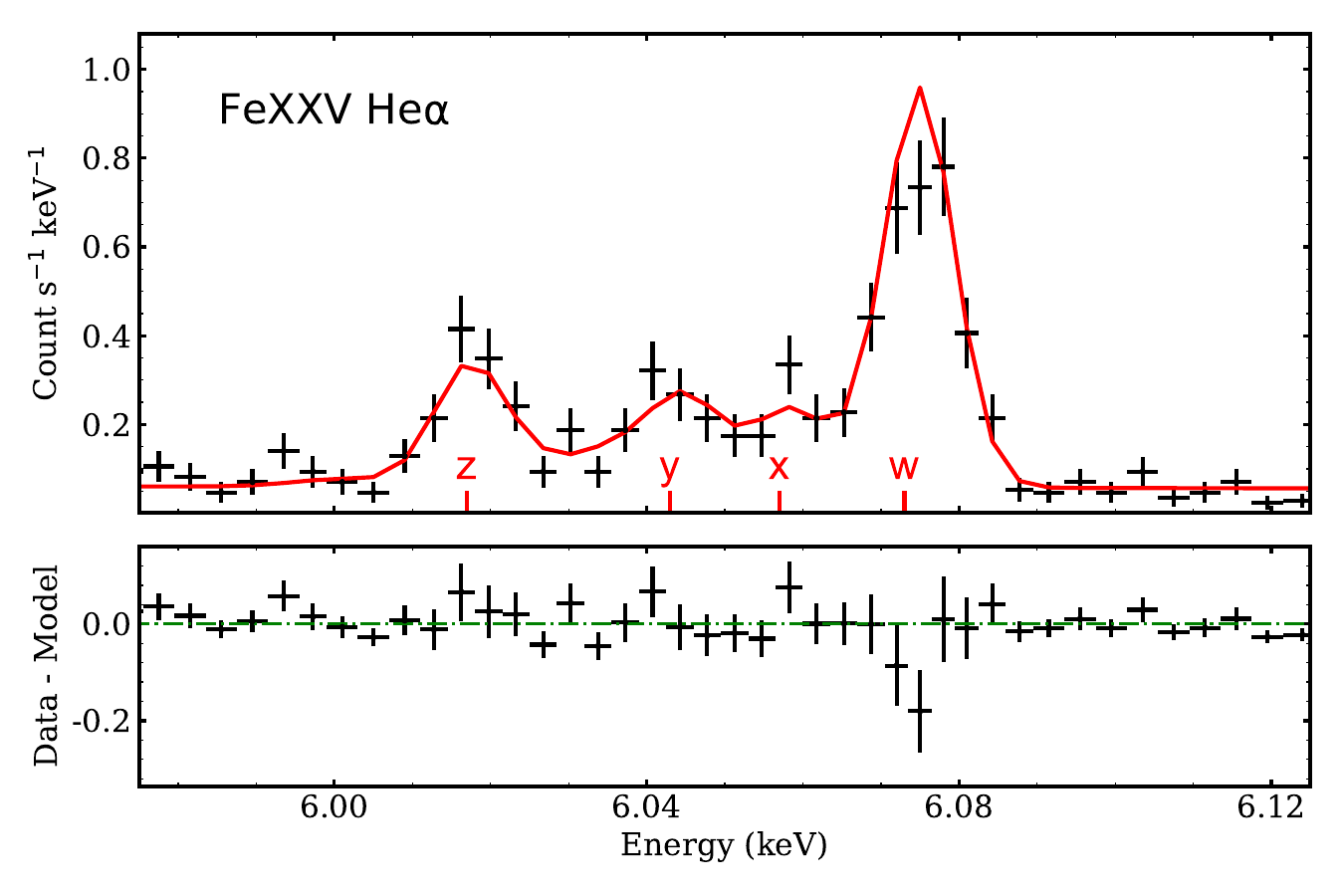}
  \end{minipage}
  \caption{Energy spectra of the entire region (23 pixels) taken during the observation period with the FW OPEN position. (Top): 2--8~keV spectrum of PKS~0745$-$191. The spectrum is fitted with \texttt{bvapec} model. (Bottom): Zoomed up in Fe \emissiontype{XXV} He$\alpha$, which consists of the resonance line 'w', intercombination lines 'x' and 'y', and forbidden line 'z'.
  {Alt text: Two line graphs showing the spectrum and fitting results. In the top panel, x axis shows the energy from 2 to 8 keV. The y axis shows the count per second and per keV. In the bottom panel, x axis shows the energy from 5.975 to 6.125 keV.}
  }
  \label{fig:all_region_spectrum}
\end{figure}

We analyzed three regions; the entire (23 pixels), the central (4 pixels), and the outer (19 pixels) region (Figure \ref{fig:image}). To evaluate the effect of RS, we performed spectral fits using two models: one with a single \texttt{bvapec} component, and the other where the w (resonance) and z (forbidden) lines of Fe \emissiontype{XXV} He$\alpha$ were excluded from \texttt{bvapec} and modeled separately with Gaussians (zgauss). The redshift of the Gaussian components are fixed to that of the \texttt{bvapec}.
The w/z ratio is a good indicator to diagnose RS, because suppression by RS depends on the oscillator strength of each line.

Each region contains mutual fluxes of each area by the point spread function (PSF) of the Resolve ($\sim 1.3^{\prime}$). 
The central region (1.0$^{\prime}$) is smaller than the PSF, allowing many photons to leak into the outer region.
This effect is called the spatial-spectral mixing (SSM) \citep{xrismcollaboration2025}.
To take into account the SSM, we fitted spectra of central and outer regions simultaneously considering the mixing components. 
To see the magnitude of the SSM effect, we present results for both the SSM-considered case (hereafter SSM analysis) and SSM-ignored case (non-SSM case; fitting each region separately without considering the mixing).
The obtained parameters in spectral fitting are summarized in Table \ref{tab:best-fit-combined}.

\begin{table*}[htbp]
  \begin{threeparttable}
  \caption{Best-fit parameters of each region spectrum for Non-SSM and SSM cases. Quoted errors are $1\sigma$.}
  \label{tab:best-fit-combined}
  \centering
  \begin{tabular*}{\textwidth}{@{\extracolsep{\fill}} l lll ll}
    \hline
    & \multicolumn{3}{c}{\textbf{Non-SSM}} & \multicolumn{2}{c}{\textbf{SSM}} \\[-1pt]
    \cline{2-4}\cline{5-6}
    Parameter & Entire & Central & Outer & Central & Outer \\ \hline\hline
    \multicolumn{6}{l}{\texttt{bvapec}} \\[2pt]
    $kT$ (keV)                & $5.936^{+0.097}_{-0.092}$ & $5.37^{+0.15}_{-0.13}$ & $6.28^{+0.13}_{-0.13}$ & $4.55^{+0.13}_{-0.13}$ & $7.85^{+0.23}_{-0.21}$ \\
    $z_{\mathrm{Open}}$      & $0.10293^{+5.2\times10^{-5}}_{-5.3\times10^{-5}}$ & $0.102854^{+5.3\times10^{-5}}_{-9.7\times10^{-5}}$ & $0.102977^{+7.1\times10^{-5}}_{-7.4\times10^{-5}}$ & $0.10270^{+7.0\times10^{-5}}_{-1.1\times10^{-4}}$ & $0.10337^{+7.0\times10^{-5}}_{-1.4\times10^{-4}}$ \\
    $z_{\mathrm{OBF}}$       & $0.10297^{+6.0\times10^{-5}}_{-1.4\times10^{-4}}$ & $0.10287^{+1.5\times10^{-4}}_{-9.0\times10^{-5}}$ & $0.10290^{+1.0\times10^{-4}}_{-1.7\times10^{-4}}$ & $0.10304^{+1.2\times10^{-4}}_{-1.6\times10^{-4}}$ & $0.10316^{+1.7\times10^{-4}}_{-2.2\times10^{-4}}$ \\
    $z_{\mathrm{ND}}$        & $0.10302^{+1.7\times10^{-4}}_{-1.3\times10^{-4}}$ & $0.10300^{+2.3\times10^{-4}}_{-2.4\times10^{-4}}$ & $0.10301^{+1.8\times10^{-4}}_{-2.7\times10^{-4}}$ & $0.10313^{+1.8\times10^{-4}}_{-2.6\times10^{-4}}$ & $0.10306^{+3.2\times10^{-4}}_{-4.3\times10^{-4}}$ \\
    $z_{\mathrm{Be}}$        & $0.103018^{+5.6\times10^{-5}}_{-6.6\times10^{-5}}$ & $0.10307^{+8.2\times10^{-5}}_{-9.8\times10^{-5}}$ & $0.102943^{+8.9\times10^{-5}}_{-8.4\times10^{-5}}$ & $0.10322^{+1.1\times10^{-4}}_{-1.1\times10^{-4}}$ & $0.10326^{+1.0\times10^{-4}}_{-1.0\times10^{-4}}$ \\
    $\sigma_{\mathrm{turb}}$ (km\,s$^{-1}$) & $152^{+10}_{-10}$ & $127^{+15}_{-18}$ & $163^{+16}_{-15}$ & $111^{+20}_{-17}$ & $97^{+22}_{-30}$ \\
    $Z_{\mathrm{Fe}}$ ($Z_{\odot}$) & $0.393^{+0.014}_{-0.014}$ & $0.406^{+0.022}_{-0.022}$ & $0.383^{+0.018}_{-0.018}$ & $0.427^{+0.024}_{-0.022}$ & $0.396^{+0.030}_{-0.026}$ \\
    $Z_{\mathrm{Ni}}$ ($Z_{\odot}$) & $0.39^{+0.11}_{-0.10}$ & $0.39^{+0.17}_{-0.15}$ & $0.38^{+0.15}_{-0.14}$ & $0.51^{+0.20}_{-0.19}$ & $0.27^{+0.21}_{-0.20}$ \\
    $\rm apec\ norm \tnote{*} $           & $0.06728^{+7.5\times10^{-4}}_{-7.6\times10^{-4}}$ & $0.0959^{+1.3\times10^{-3}}_{-1.3\times10^{-3}}$ & $0.05735^{+6.7\times10^{-4}}_{-6.2\times10^{-4}}$ & $0.03267^{+6.6\times10^{-4}}_{-6.7\times10^{-4}}$ & $0.02127^{+4.0\times10^{-4}}_{-4.0\times10^{-4}}$ \\
    cstat/d.o.f              & 5717.68/6224 & 3246.20/4022 & 4505.93/5170 & \multicolumn{2}{c}{7739.88/9184} \\
    \hline
    \multicolumn{6}{l}{\texttt{bvapec (excluding w,z) + zgauss(w) + zgauss(z)}} \\[2pt]
    $kT$ (keV)                & $5.866^{+0.098}_{-0.097}$ & $5.33^{+0.14}_{-0.12}$ & $6.27^{+0.14}_{-0.14}$ & $4.59^{+0.12}_{-0.13}$ & $7.77^{+0.23}_{-0.21}$ \\
    $z_{\mathrm{Open}}$      & $0.102944^{+4.6\times10^{-5}}_{-5.8\times10^{-5}}$ & $0.10284^{+5.0\times10^{-5}}_{-1.0\times10^{-4}}$ & $0.102998^{+6.0\times10^{-5}}_{-8.6\times10^{-5}}$ & $0.10276^{+5.0\times10^{-5}}_{-1.3\times10^{-4}}$ & $0.10337^{+8.0\times10^{-5}}_{-1.0\times10^{-4}}$ \\
    $z_{\mathrm{OBF}}$       & $0.102914^{+8.9\times10^{-5}}_{-9.7\times10^{-5}}$ & $0.10286^{+1.3\times10^{-4}}_{-1.1\times10^{-4}}$ & $0.10290^{+1.3\times10^{-4}}_{-1.3\times10^{-4}}$ & $0.10296^{+1.6\times10^{-4}}_{-1.1\times10^{-4}}$ & $0.10317^{+2.4\times10^{-4}}_{-1.7\times10^{-4}}$ \\
    $z_{\mathrm{ND}}$        & $0.10302^{+1.7\times10^{-4}}_{-1.2\times10^{-4}}$ & $0.10300^{+2.1\times10^{-4}}_{-2.3\times10^{-4}}$ & $0.10300^{+1.9\times10^{-4}}_{-2.6\times10^{-4}}$ & $0.10309^{+2.0\times10^{-4}}_{-2.1\times10^{-4}}$ & $0.10318^{+4.0\times10^{-4}}_{-4.8\times10^{-4}}$ \\
    $z_{\mathrm{Be}}$        & $0.103001^{+5.5\times10^{-5}}_{-6.6\times10^{-5}}$ & $0.103038^{+9.5\times10^{-5}}_{-9.0\times10^{-5}}$ & $0.10291^{+1.1\times10^{-4}}_{-6.0\times10^{-5}}$ & $0.10319^{+1.0\times10^{-4}}_{-1.1\times10^{-4}}$ & $0.10283^{+1.3\times10^{-4}}_{-1.7\times10^{-4}}$ \\
    $\sigma_{\mathrm{turb}}$ (km\,s$^{-1}$) & $121^{+17}_{-17}$ & $120^{+29}_{-30}$ & $124^{+24}_{-22}$ & $78^{+34}_{-49}$ & $110^{+33}_{-41}$ \\
    $Z_{\mathrm{Fe}}$ ($Z_{\odot}$) & $0.422^{+0.021}_{-0.021}$ & $0.44^{+0.034}_{-0.034}$ & $0.394^{+0.025}_{-0.023}$ & $0.462^{+0.034}_{-0.033}$ & $0.388^{+0.037}_{-0.035}$ \\
    $Z_{\mathrm{Ni}}$ ($Z_{\odot}$) & $0.40^{+0.11}_{-0.10}$ & $0.38^{+0.17}_{-0.15}$ & $0.37^{+0.14}_{-0.13}$ & $0.51^{+0.21}_{-0.19}$ & $0.26^{+0.22}_{-0.21}$ \\
    apec norm$^{*}$           & $0.06777^{+7.3\times10^{-4}}_{-6.4\times10^{-4}}$ & $0.0962^{+2.3\times10^{-4}}_{-3.0\times10^{-4}}$ & $0.05766^{+7.6\times10^{-4}}_{-3.4\times10^{-4}}$ & $0.03221^{+7.0\times10^{-4}}_{-6.8\times10^{-4}}$ & $0.02150^{+4.1\times10^{-4}}_{-4.0\times10^{-4}}$ \\
    $\mathrm{norm}_w^{\dagger}$ ($10^{-5}$) & $7.56^{+0.44}_{-0.41}$ & $10.65^{+0.97}_{-0.90}$ & $6.46^{+0.48}_{-0.47}$ & $3.74^{+0.34}_{-0.33}$ & $2.21^{+0.32}_{-0.27}$ \\
  $\sigma_w$ (eV [km\,s$^{-1}$]) 
& $4.64^{+0.32}_{-0.31}$ (229$^{+16}_{-15}$) 
& $3.84^{+0.46}_{-0.44}$ (190$^{+23}_{-22}$) 
& $5.06^{+0.54}_{-0.40}$ (250$^{+27}_{-20}$) 
& $3.35^{+0.38}_{-0.46}$ (165$^{+19}_{-23}$) 
& $6.11^{+0.32}_{-0.27}$ (302$^{+16}_{-13}$) \\
    $\mathrm{norm}_z^{\dagger}$ ($10^{-5}$) & $2.63^{+0.28}_{-0.27}$ & $4.28^{+0.66}_{-0.59}$ & $2.00^{+0.30}_{-0.27}$ & $1.62^{+0.20}_{-0.23}$ & $0.45^{+0.16}_{-0.14}$ \\

$\sigma_z$ (eV [km\,s$^{-1}$]) 
& $3.28^{+0.54}_{-0.48}$ (162$^{+27}_{-24}$) 
& $2.57^{+0.59}_{-0.78}$ (127$^{+29}_{-39}$) 
& $3.32^{+0.82}_{-0.84}$ (164$^{+41}_{-41}$) 
& $2.37^{+0.68}_{-0.58}$ (116$^{+34}_{-29}$) 
& $0.90^{+2.00}_{-0.90}$ (44$^{+99}_{-44}$) \\
    w/z ratio               & $2.87^{+0.34}_{-0.34}$ & $2.49^{+0.41}_{-0.43}$ & $3.23^{+0.49}_{-0.53}$ & $2.31^{+0.39}_{-0.35}$ & $4.9^{+1.7}_{-1.8}$ \\
    optically thin            & 3.25 & 3.17 & 3.31 & 3.08 & 3.47 \\
    cstat/d.o.f               & 5706.57/6220 & 3238.72/4018 & 4497.52/5166 & \multicolumn{2}{c}{7725.53/9176} \\

    \hline
  \end{tabular*}
  \begin{tablenotes}
  \item[*] Normalization of the \texttt{bvapec} model. Defined by $\frac{10^{-14}}{4\pi [D_A(1+z)]^2 }\int n_e n_H dV$, where $D_A$ is the angular diameter distance.
  \item[$\dagger$] Integrated flux of the \texttt{zgauss} model.
  \end{tablenotes}
  \end{threeparttable}
\end{table*}

\subsection{Non-SSM analysis}
First we fitted the spectrum of the entire region with \texttt{bvapec} model (Figure \ref{fig:all_region_spectrum}). 
The Fe \emissiontype{XXV} He$\alpha$ w line intensity is overpredicted by the optically thin, single-temperature plasma model at collisional ionization equilibrium (CIE) (Figure \ref{fig:all_region_spectrum} bottom). The obtained temperature is $kT = 5.936^{+0.097}_{-0.092}$ keV, and the turbulent velocity is $152\pm 10$ km s$^{-1}$.
A suppression of the w intensity is typically caused by RS. 
We show the fitting result with \texttt{bvapec(excluding w,z) + zgauss(w) + zgauss(z)} of each region in table \ref{tab:best-fit-combined}.
The w/z flux ratio of the entire region is $2.88^{+0.31}_{-0.28}$ (table \ref{tab:best-fit-combined}) is smaller than the optically thin ratio of 3.25 (at 5.866~keV; table~\ref{tab:best-fit-combined}).
The w/z flux ratio in the central region and outer region are measured to be $2.49^{+0.41}_{-0.43}$ and $3.23^{+0.49}_{-0.53}$, respectively corresponding to 78$^{+12}_{-14}\%$ and 98$^{+15}_{-16}\%$ of those of optically thin cases.
These radial trend is consistent with expectations from RS, where the higher ion density in the core leads to stronger suppression. 
RS reduces the emission line intensity and broadens the line \citep{hitomicollaboration2018}. 
In the central region, the emission line broadenings of the z and w lines are $2.57^{+0.59}_{-0.43}$ eV and $3.84^{+0.46}_{-0.44}$ eV, respectively. 
The corresponding turbulent velocities at 5.33~keV (table \ref{tab:best-fit-combined}) are $\rm 66^{+24}_{-22}\ km\ s^{-1}$ and $\rm 143^{+39}_{-30}\ km\ s^{-1}$, with the w line being significantly broader.

In the entire region, the difference in the redshift between different FW datasets is $2.7\times 10^{-4}$ (0.54 eV at 6~keV) and this is comparable to the gain uncertainty of $\sim 0.6$ eV at 5.4--8.0~keV (Appendix 1).
The redshift of FW OPEN after heliocentric correction is $z=0.10286\pm 0.00012$, which includes the gain uncertainty.

\subsection{SSM analysis}
To take into account the SSM, we fitted the spectra of the central and outer region simultaneously.
The leakage from one region to another was estimated by calculating corresponding ARFs; e.g., for the leakage from central region to outer region, we simulated the number fraction of photons emitted from the projected sky corresponding to the central region and detected by the pixels in the outer region. 
The source spectrum of the projected sky area corresponding to the 12 pixels that were not used in the analysis was assumed to be identical to that of the outer region.
The leakage from outside the Resolve FoV was ignored because it accounted for less than 3\% of the outer-region flux. 
The leakage from the outer to the central region is about 20\% (at 6~keV) of the total photon counts, 
while that from the central to the outer region is about 50\% (at 6~keV) of the total photon counts.

\begin{figure*}[htbp]
  \centering
  \begin{minipage}[b]{0.48\textwidth}
    \centering
    \includegraphics[width=\linewidth]{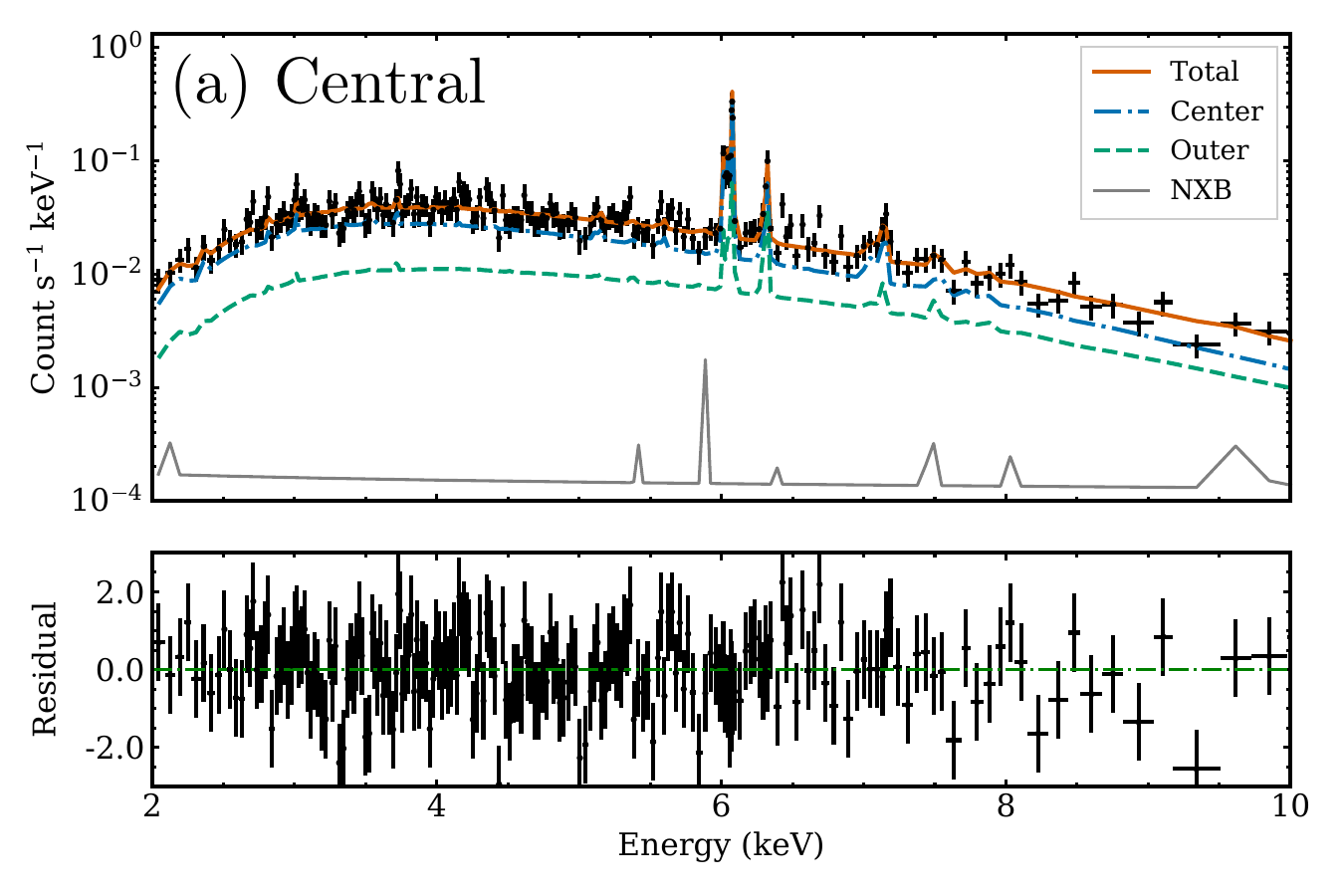}
  \end{minipage}
  \hfill
  \begin{minipage}[b]{0.48\textwidth}
    \centering
    \includegraphics[width=\linewidth]{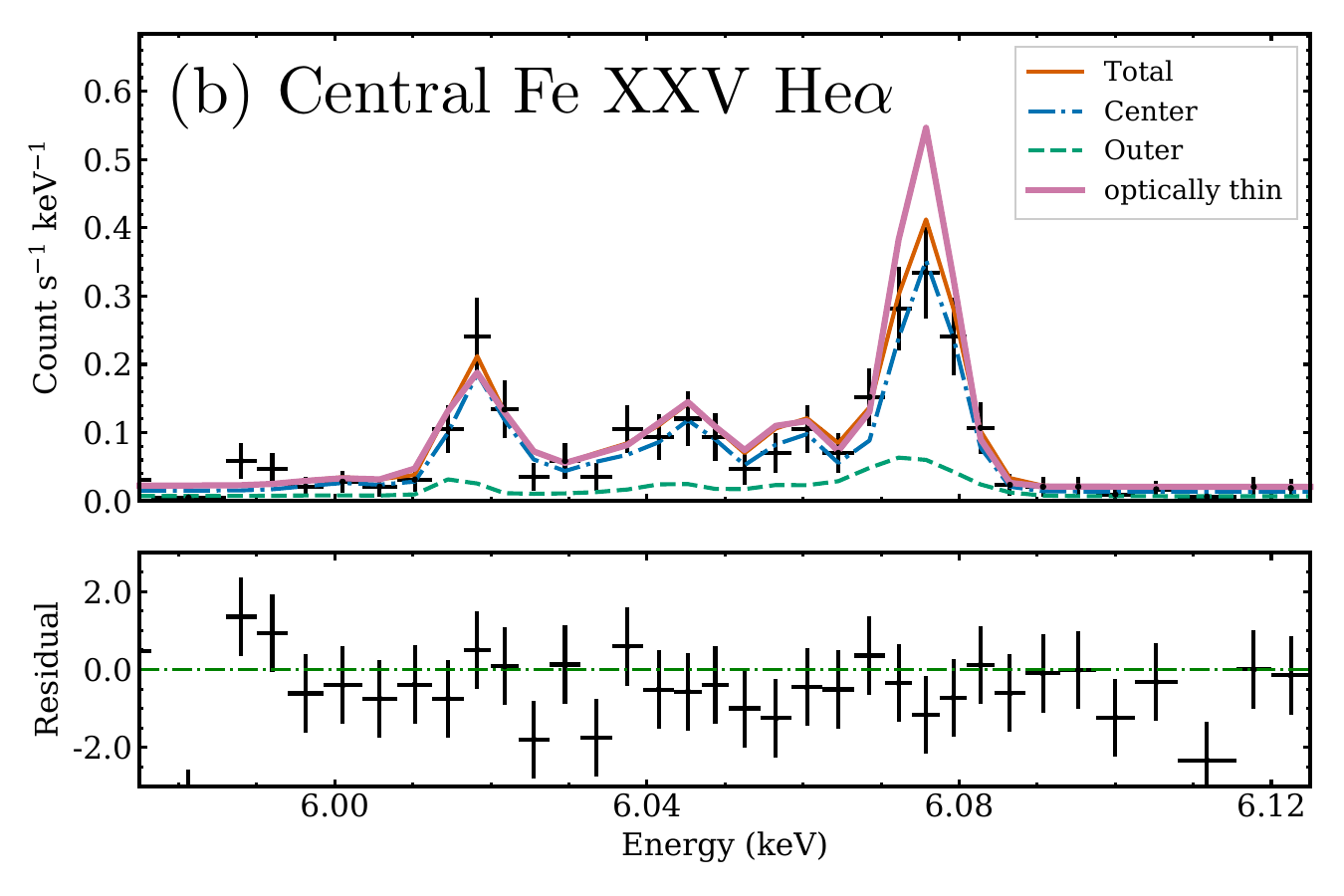}
  \end{minipage}

  \begin{minipage}[b]{0.48\textwidth}
    \centering
    \includegraphics[width=\linewidth]{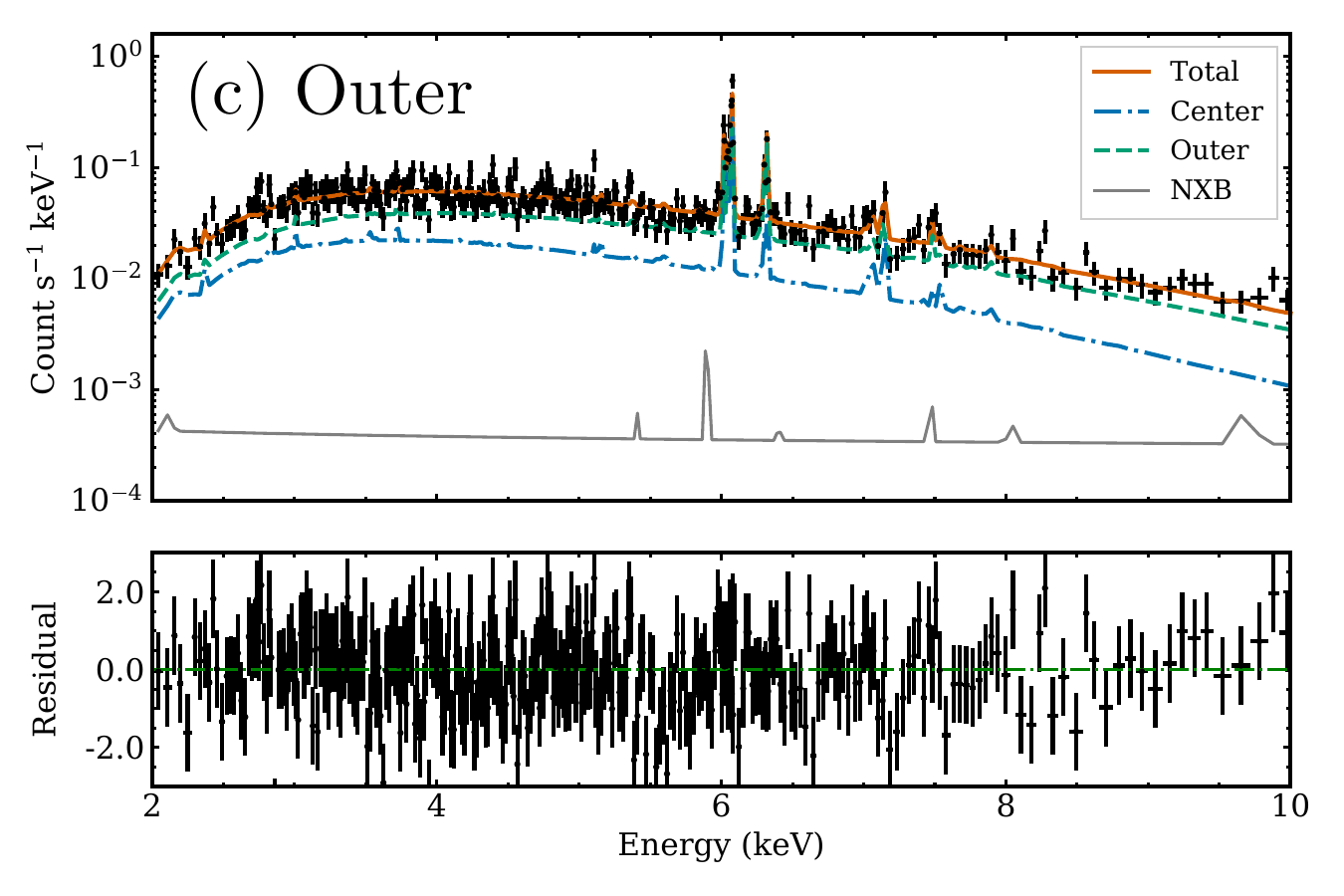}
  \end{minipage}
  \hfill
  \begin{minipage}[b]{0.48\textwidth}
    \centering
    \includegraphics[width=\linewidth]{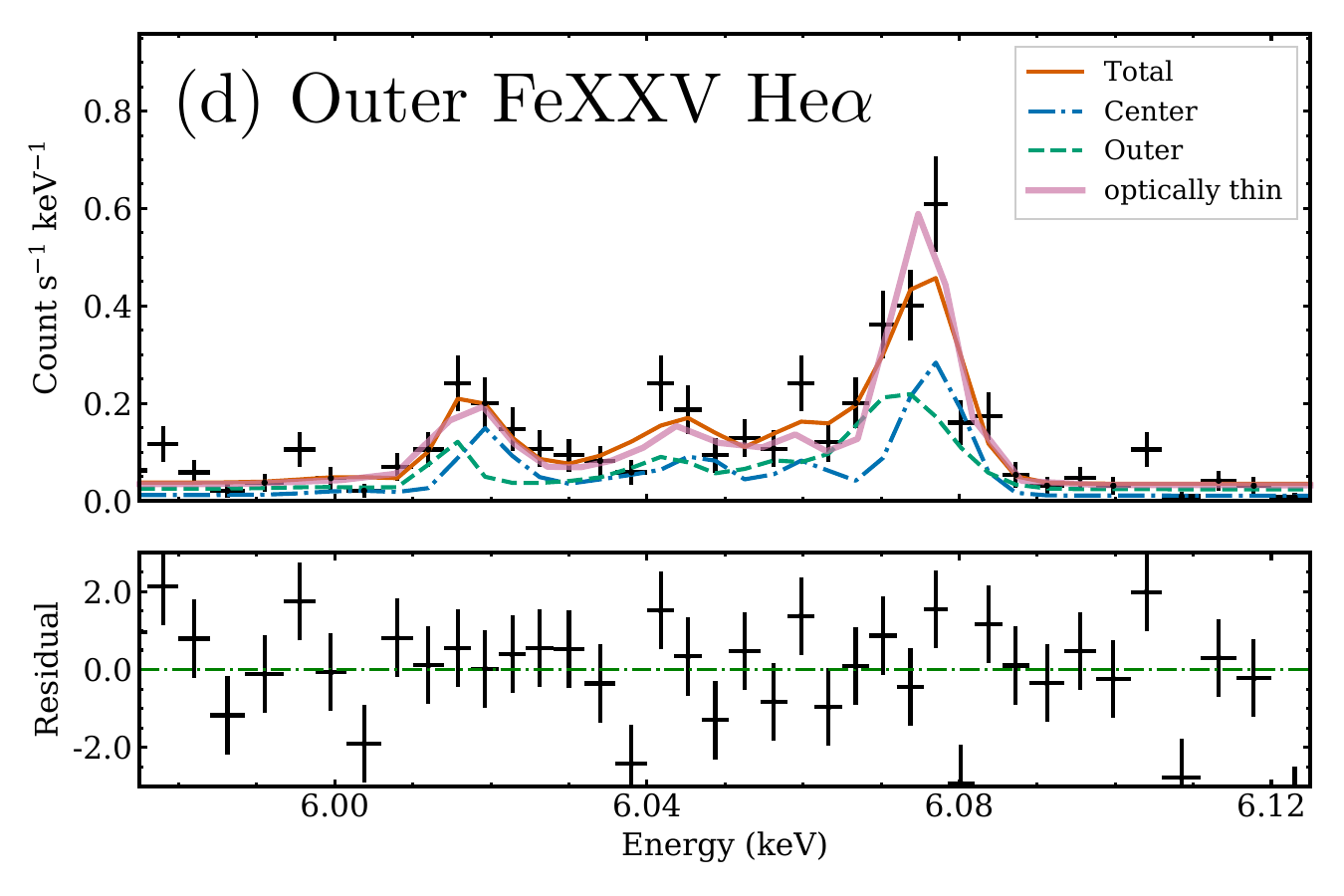}
  \end{minipage}

  \caption{Energy spectra of central and outer regions observed with \textit{Resolve}. (a) 2--10~keV spectrum of the central region (central four pixels). (b) 5.975--6.125~keV spectrum of the central region covering the Fe \emissiontype{XXV} He$\rm \alpha$ line. (c) 2--10~keV spectrum of the outer region (outer 19 pixels). (d) 5.975--6.125~keV spectrum of the outer region. The orange solid line is the best-fitting results of \texttt{bvapec+zgauss(w)+zgauss(z)} model. The blue dash-dotted, green dashed, and gray solid lines show the components from the central, outer, and NXB model, respectively. The magenta solid lines are the \texttt{bvapec} model of $kT=5.33 \ \rm keV$ (central) and $6.27 \ \rm keV$ (outer). 
  {Alt text: Four line graphs showing the spectrum and best-fitting results with model components. The y axis is the count per second and per keV.}
  }
  \label{fig:center_outer_spectrum}
\end{figure*}

We showed the spectrum of the central and the outer region in Figure \ref{fig:center_outer_spectrum}.
The central and outer temperatures are $4.55^{+0.13}_{-0.13}$ keV and $7.85^{+0.23}_{-0.21}$ keV, respectively, for the single \texttt{bvapec} model. 
The larger separation between the two temperature compared to the non-SSM analysis clearly shows that the SSM effect is significant. The turbulent velocity of each region is $78^{+34}_{-49} \rm {km\ s^{-1}}$ and $110^{+33}_{-41} \rm {km\ s^{-1}}$, respectively. 

The w/z ratios of $2.31^{+0.29}_{-0.35}$ and $4.9^{+1.7}_{-1.8}$ were obtained for the central and outer regions, respectively.
The w/z ratio of the central region is more suppressed than that of the optically thin case as in the non-SSM analysis.
On the other hand, the outer region of the w/z ratio is consistent with the optically thin case; it is slightly higher, but the difference is statistically marginal ($<1\sigma$). As RS scattered photons at the center of the cluster of galaxies, while outside the cluster the intensity can increase \citep{gilfanov1987}.

The average redshift over all FW datasets is $z = 0.10292 \pm 0.00022$ in the central region and $z = 0.10305 \pm 0.00027$ in the outer region.
Taking the difference between FW datasets as systematic uncertainty, the upper limit on the redshift difference between the two regions is $\Delta z < 6.3 \times 10^{-4}$, corresponding to $v_{\rm bulk} < 190\ \rm km\ s^{-1}$.

Both analysis method of the non-SSM and SSM cases showed that the w/z flux ratio of the central region is suppressed by RS. 
Systematic uncertainties arise from both the plasma codes and astrophysical modeling such as the multi-temperature modeling of the ICM. The magnitude of the plasma code uncertainty can be gauged by comparing differences between different codes or versions. These are discussed in Appendices 3 and 4, respectively, and they have little impact on the results.

\section{Radiative Transfer Simulation of the Resonant Scattering}
\label{sec:RS_Sim}
\subsection{Construction of the RS simulation}
\begin{figure*}
\centering
\begin{minipage}[htbp]{0.48\textwidth}
\centering
\includegraphics[width=\textwidth]{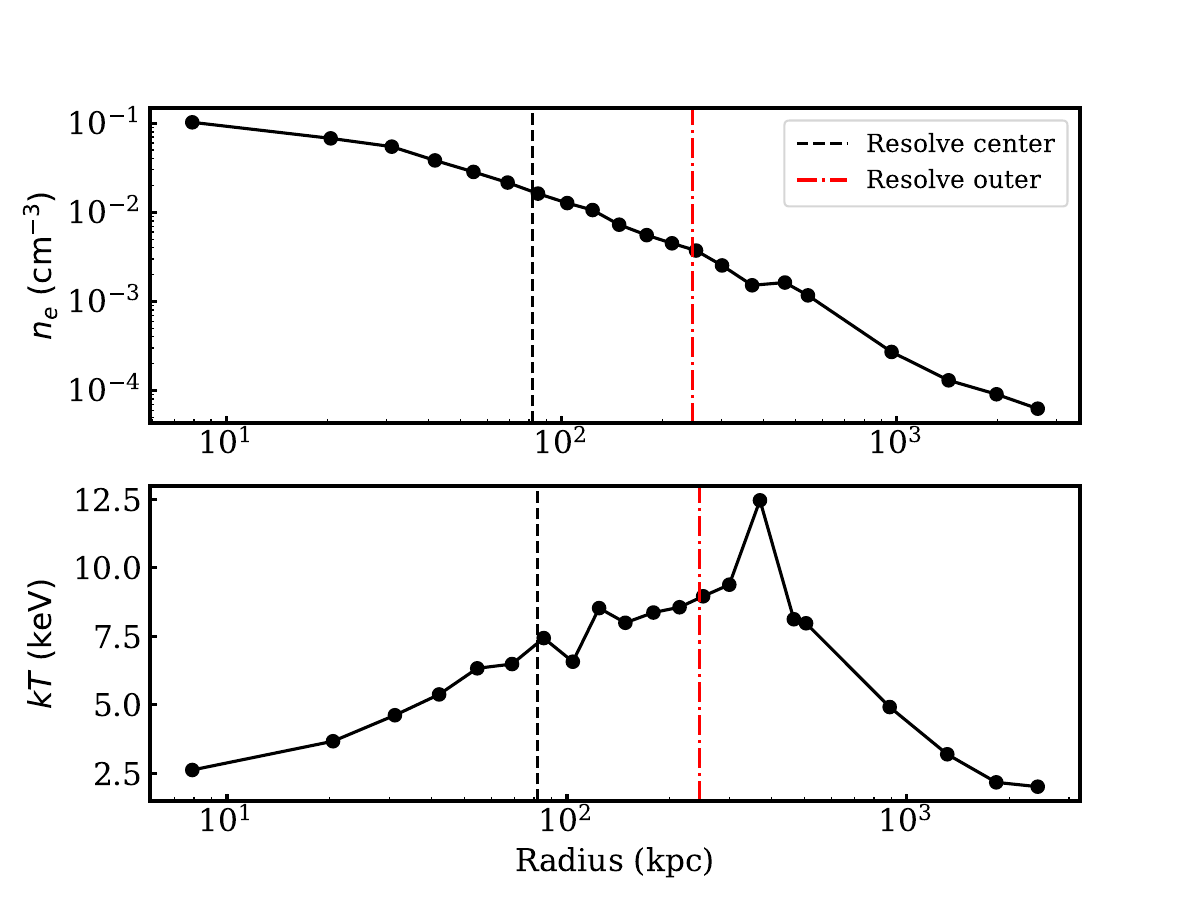}
\end{minipage}
\begin{minipage}[htbp]{0.48\textwidth}
\centering
\includegraphics[width=\textwidth]{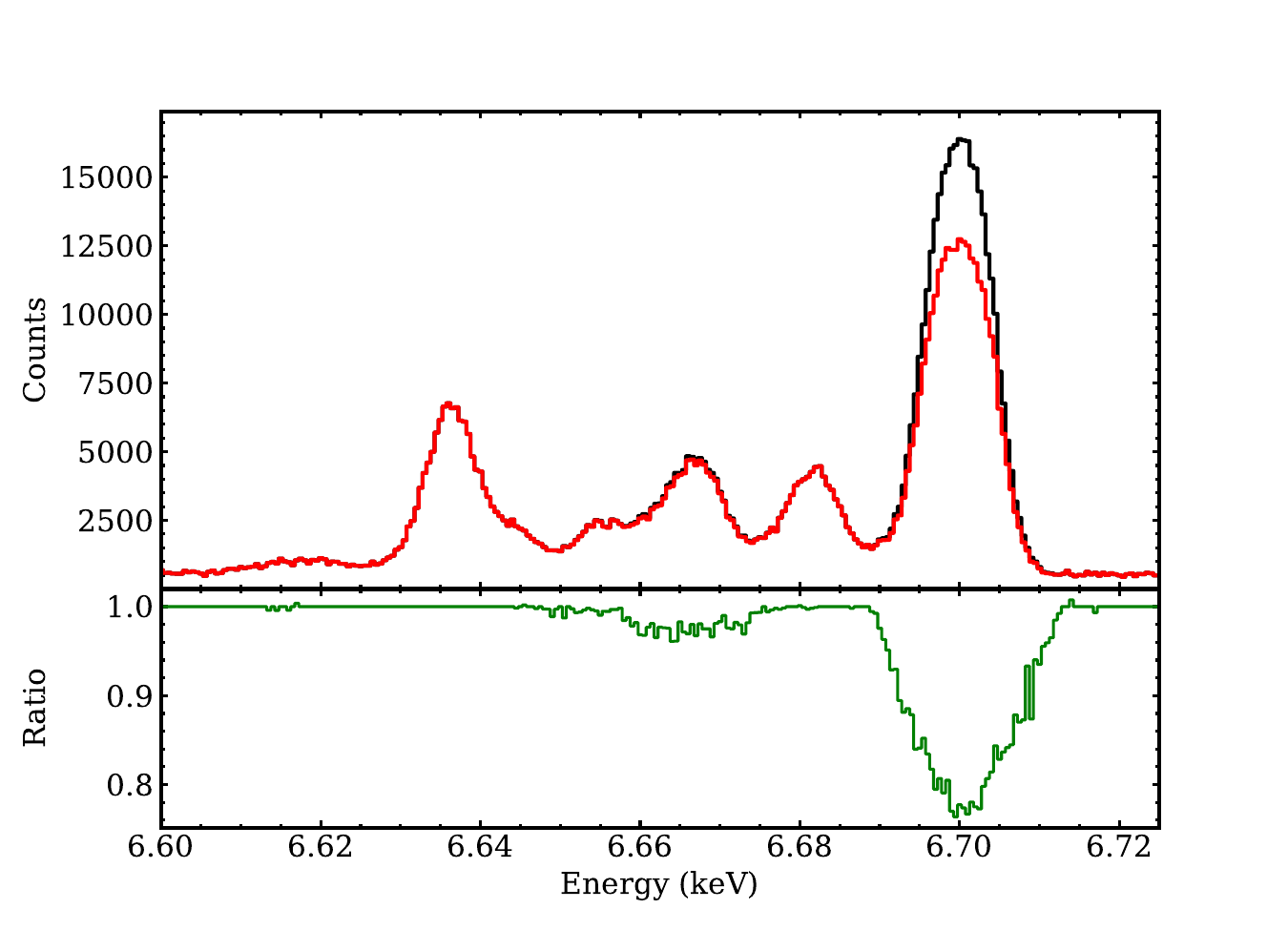}
\end{minipage}
\caption{(Left): The setting parameters of the RS simulation. Up to 500 kpc radius we used the results from \citet{sanders2014} and up to 1000 kpc we used the results from \citep{walker2012}. The metal abundance was constant at 0.4 Solar value of \citet{anders1989} for \citet{sanders2014} and \citet{walker2012}. (Right): The simulated spectrum of the turbulent velocity is 100 $\rm km\ s^{-1}$ in PKS~0745$-$191 central region ($r=0.5^{\prime}$). The black line shows without RS case (optically thin) and the red line shows with RS. The bottom panel shows count ratio of with and without the RS effect at $\sigma_{\rm turb}=100 \rm \ km \ s^{-1}$.
{Alt text: Three line graphs. In the left panel, the radial profile of the electro density and temperature are shown. The x axis shows radius from 6 kilo parsec to 3000 kilo parsec. In the right panel, the simulated spectrum of with and without RS are shown. The x axis shows the energy from 6.5 keV to 6.725 keV. The y axis shows the count. The bottom panel shows the count ratio of with and without the RS effect.}
}
\label{fig:RS_Sim_setting}
\end{figure*}

When combined with simulations, the w/z line ratio provides a powerful diagnostic of turbulent velocity and bulk motion on small scales within the ICM.
We estimate the velocity field of PKS~0745$-$191 by constructing a radiative transfer simulation including RS using the Monte Carlo method and obtaining w/z ratios for various velocity structures.
We assumed a spherically symmetric ICM profile with the electron density and temperature obtained by Chandra \citep{sanders2014} and Suzaku \citep{walker2012} observations.
The metal abundance was fixed at $0.4 \ Z_\odot$ at all radii, consistent with Resolve observations, which show no significant difference in Fe abundance between the central and outer regions.
The ICM was divided into 400 spherical shells following the profiles shown in Figure \ref{fig:RS_Sim_setting}.
Initial photon positions were sampled according to an $n_e^2$ distribution, and the initial photon energies were sampled from an APEC model (\textsc{AtomDB} v3.0.9) including both lines and continuum emissivities.
The total line width was taken as the quadratic sum of thermal and turbulent broadening.

The cross section of the RS is given by \citep{shigeyama1998}
\begin{equation}
\sigma_{\rm RS}(E) = \frac{\pi h r_e c f}{\sqrt{2\pi}\sigma}\exp\left[-\frac{(E-E_0)^2}{2\sigma^2} \right]
\end{equation}
where $h$ is the Planck constant, $r_e$ is the classical electron radius, $c$ is the speed of light, and $f$ is the oscillator strength of the transition.
The Doppler width $\sigma$ due to thermal and turbulent motions is
\begin{equation}
\sigma = E_0 \left[\frac{kT_{\rm ion}}{Am_pc^2}+\frac{\sigma_{\rm turb}^2}{c^2}\right]^{1/2}
\end{equation}
where $T_{\rm ion}$ is the ion temperature, we assume the ion temperature is equal to the electron temperature, $A$ the atomic mass in units of proton mass, $m_p$ the proton mass, and $\sigma_{\rm turb}$ the one-dimensional turbulent velocity.
The corresponding optical depth of the transition is
\begin{equation}
\tau_{\rm RS}(E) = \int \sigma_{\rm RS}(E) n_p Z \delta_i dl
\end{equation}
where $n_p$ is the proton density, $Z$ the metal abundance, and $\delta_i$ the ion fraction.
Table \ref{tab:tau_atomic_data} summarizes the relevant line parameters and optical depths (for zero $\sigma_{\rm turb}$ and $v_{\rm bulk}$), based on \textsc{AtomDB} v3.0.9. 

At each simulation step, the photon path length $s$ is sampled from an exponential distribution via inverse-transform sampling,
$s = -\ln\xi , / \Sigma$, where $\xi \sim \mathcal{U}[0,1]$, $\mathcal{U}$ is the uniform distribution, and $\Sigma$ is the total macroscopic cross section; the photon is then advanced by this distance to its next interaction point. 
Scattering events were iterated until the photon reached a radius of $r=1$ Mpc, corresponding to $\sim 3/4\times r_{500}$ of PKS~0745$-$191.
Scattered photons were assumed to be re-emitted isotropically.
The post-scattering photon energy was computed as
\begin{equation}
\label{eq:RS_energy}
E = E_0 \left(1+\frac{\textbf v_{\rm ion}\cdot \textbf m}{c}\right)
\end{equation}
where $\textbf{v}_{\rm ion}$ is the ion velocity, $\textbf{m}$ the photon propagation direction, and $E_0$ the line energy in rest frame.
Ion velocities participating in scattering were sampled from the conditional probability
\begin{equation}
\label{eq:RS_velocity}
f(E, v_{\rm ion}) = \frac{f_{\rm MB}(v)\tau_{\rm ion}(E, v)}{\int f_{\rm MB}(v)\tau_{\rm ion}(E, u)du} = \frac{f_{\rm MB}(v)\tau_{\rm ion}(E, v)}{\tau_{\rm RS}}
\end{equation}
where $f_{\rm MB}(v)$ is the Maxwell-Boltzmann distribution, and $\tau_{\rm ion}$ is the RS cross section in the ion frame. When a photon energy $E$ causes RS, the ion velocity is sampled according to equation (\ref{eq:RS_velocity}), and the post-scattering energy is determined by equation (\ref{eq:RS_energy}).

Bulk motion is defined as translational motion relative to the observer, and the energy of a photon is calculated as $E=E_0\left(1+\frac{v_{\rm bulk}}{c}\right)$.
\begin{table*}[t]
\centering
\caption{Table of line transitions with oscillator strengths and optical depths in the PKS 0745-191. The optical depth is calculated at the line center energy, assuming zero turbulent and bulk velocities. The central energy and oscillator strength are from \textsc{AtomDB} v3.0.9 \citep{foster2012,foster2020}. The line IDs are defined by \citep{gabriel1972}.}
\label{tab:tau_atomic_data}
\begin{tabular}{lllllll}
\toprule
Line ID & Ion & Energy (eV) & Transition & Type & Oscillator strength $f$ & Optical depth $\tau$ \\
\midrule
u & Fe XXIV & 6616.73 & $1s^2 2s^2 \rightarrow 1s 2s 2p$ & satellite & $1.63\times10^{-2}$ & $1.10\times10^{-2}$ \\
z & Fe XXV  & 6636.58 & $1s^2~^1S_0 \rightarrow 1s 2s~^3S_1$ & forbidden & $3.03\times10^{-7}$ & $1.25\times10^{-6}$ \\
r & Fe XXIV & 6653.30 & $1s^2 2s^2 \rightarrow 1s 2s 2p$ & satellite & $1.57\times10^{-1}$ & $1.05\times10^{-1}$ \\
q & Fe XXIV & 6661.88 & $1s^2 2s^2 \rightarrow 1s 2p$ & satellite & $4.89\times10^{-1}$ & $3.27\times10^{-1}$ \\
y & Fe XXV  & 6667.55 & $1s^2~^1S_0 \rightarrow 1s 2p~^3P_1$ & intercomb. & $5.79\times10^{-2}$ & $2.38\times10^{-1}$ \\
t & Fe XXIV & 6676.59 & $1s^2 2s^2 \rightarrow 1s 2p$ & satellite & $9.62\times10^{-3}$ & $6.46\times10^{-2}$ \\
x & Fe XXV  & 6682.30 & $1s^2~^1S_0 \rightarrow 1s 2p~^3P_2$ & intercomb. & $1.70\times10^{-5}$ & $6.96\times10^{-5}$ \\
w & Fe XXV  & 6700.40 & $1s^2~^1S_0 \rightarrow 1s 2p~^1P_1$ & resonance & $7.19\times10^{-1}$ & $2.92$ \\
\bottomrule
\end{tabular}
\end{table*}
A total of $10^7$ photons were simulated in the 6.5-6.8 keV band, using the same spectral model as adopted for the observational fitting.
To verify the accuracy of the simulation framework, we applied it to the Hitomi observation of the Perseus cluster and reproduced the published results of \citet{hitomicollaboration2018}.

\begin{figure*}[htbp]
  \centering
    \includegraphics[width=\linewidth]{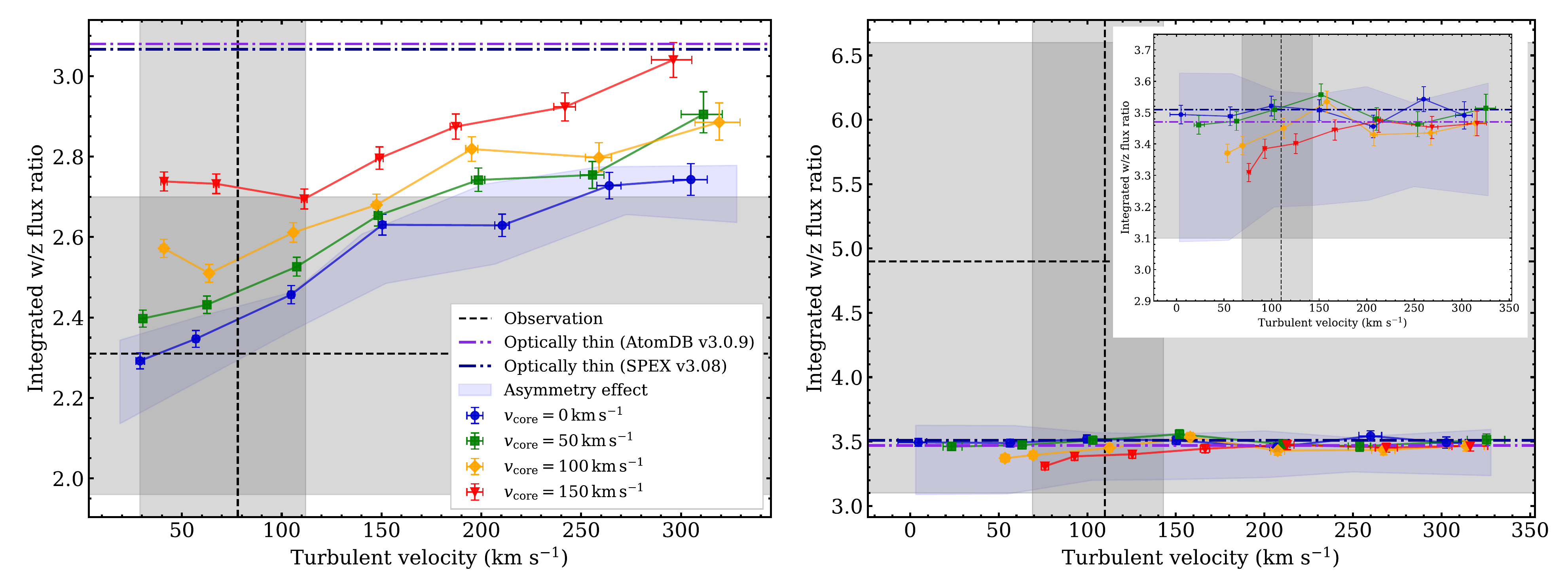}
  \caption{Integrated w/z flux ratio of the RS simulation result as a function of turbulent velocity and $v_{\rm core}$ for each region. The blue, green, yellow and red points correspond to $v_{\rm core}=0,50,100,150$ $\rm km\ s^{-1}$. The black dashed lines are the measured w/z flux ratio and turbulent velocity obtained by \texttt{bvapec + zgauss(w) + zgauss(z)} model. The gray spans and error bars are the 1$\sigma$ uncertainty. The w/z flux ratio of the optically thin case is shown by the purple (\textsc{AtomDB} v3.0.9) \citep{foster2012} and the dark blue (\textsc{SPEX} v3.08) \citep{kaastra1996} dashed dot lines. The blue spans are simulation result with $v_{\rm core}=0\ \rm km\ s^{-1}$ which used east and west profile of the \citet{sanders2014}. 
  (Left): The simulation result in the central region (Right): The simulation result in the outer region. 
  {Alt text: Two line graphs showing the comparison of w/z flux ratio as a function of turbulent velocity and bulk velocity of the core. The x axis shows turbulent velocity.}
  }
  \label{fig:wz_fourpanel}
\end{figure*}

\subsection{RS simulation result}
Figure~\ref{fig:RS_Sim_setting} right shows the simulated spectrum for $\sigma_{\rm turb}=\rm 100\ km \ s^{-1}$ in the central region of PKS~0745$-$191 ($r=0.5^{\prime}$).
As anticipated from Table~\ref{tab:tau_atomic_data}, RS strongly suppresses the He$\alpha$ w line and also weakens He$\alpha$ $y$.
We first consider a case in which the turbulent velocity is spatially constant and the bulk velocity is zero. 
Results for $\sigma_{\rm turb}=0$--$300 \ \rm km \ s^{-1}$ are summarized in Figure~\ref{fig:wz_fourpanel} (blue points). We focus here on the SSM case; for the non-SSM case, the simulation is described in Appendix 5. 

Next, we explore cases with constant $\sigma_{\rm turb}$ and non\mbox{-}zero bulk velocities. 
PKS~0745$-$191 has a cool\mbox{-}core, which we define as the radius where the radiative cooling time falls below $7.7$~Gyr (evaluated at $z=1$), corresponding to $r_{\rm cc}\simeq115$~kpc \citep{sanders2014}.
In order to treat photons spherically symmetrically in the simulation, we give a uniform bulk velocity, $v_{\rm core}$, in the radial direction from the cluster center. Such a diverging motion with a constant radial velocity is a rather artificial and non-physical process, and this is different from the bulk velocity inferred from the observations, but it allows us to test the radial bulk-velocity dependence of RS.

Since the RS probability and line width depend only on the absolute value of the bulk velocity, we simulate it in the direction that keeps the velocity positive toward the observer.
Bulk motion causes a line shift and reduces the RS probability. We adopt $v_{\rm core} = 50\text{--}150 \ \rm km\ s^{-1}$ in the simulations while $\sigma_{\rm turb}$ is spatially constant. The corresponding results are shown in Figure~\ref{fig:wz_fourpanel}.

\subsection{The effects of ICM asymmetry}
Chandra observations indicate that the electron density and temperature of PKS~0745$-$191 exhibit asymmetric profiles \citep{sanders2014}.
Although we assume spherical symmetry and use the average profile over the entire region, this assumption introduces a systematic uncertainty in the w/z ratio. 
In particular, the western and eastern regions show higher and lower electron densities, respectively, compared to the $\beta$-model \citep{sanders2014}.
To estimate this effect, we performed simulations incorporating the radial dependence of the electron density and temperature in the western and eastern regions, and examined the resulting changes in the w/z ratio.
In Figure~\ref{fig:wz_fourpanel}, the blue band indicates the simulated w/z ratio at $v_{\rm core}=0~\mathrm{km\ s^{-1}}$ for the west and east regions of PKS~0745$-$191.
The upper and lower bounds of the blue band correspond to the east and west regions, respectively.
At all turbulent velocities, the w/z ratio fluctuates by approximately $\pm0.1$. 
Because the eastern and western regions show the largest differences in electron density, the blue area in the figure can be regarded as representing the maximum systematic uncertainty due to the assumed cluster structure.

\section{Discussion}
\label{sec:discussion}

\subsection{Redshift of the ICM}
\label{sec:redshift}
We examine the redshift of the ICM measured in the X-ray and optical bands.
Table~\ref{tab:best-fit-combined} summarizes the ICM redshifts in the entire, central and outer regions. As shown in Section \ref{sec:spec}, The redshift of the entire region in FW Open is $z=0.10286\pm 0.00012$. 
\citet{hunstead1978} reported an optical redshift of $z = 0.1028 \pm 0.0005$ for PKS~0745$-$191, based on several emission lines including [O,\textsc{ii}], H$\beta$, [O,\textsc{iii}], H$\alpha$, [N,\textsc{ii}], and [S,\textsc{ii}].
The X-ray measurement of the ICM in each region is consistent with this optical determination. \citet{gingras2024} reported an optical redshift of $z = 0.1024$ for PKS~0745$-$191, based on the stellar continuum fitting.
This corresponds to a difference of 103 $\rm km | s^{-1}$ in the X-ray redshift, but it is unclear whether this is significant within the uncertainty range.

The redshift of the central region is $z = 0.10292 \pm 0.00022$ and outer region is $z = 0.10305 \pm 0.00027$.
Even after accounting for the gain uncertainty, the upper limit on the bulk velocity in the central region remains below $190\ \rm km\ s^{-1}$, indicating that PKS~0745$-$191 is dynamically quiescent along the line of sight.

\subsection{Velocity structure}
PKS~0745$-$191 has strong cool core and there are several pieces of evidence of the AGN feedback (X-ray cavities) \citep{sanders2014}.
In sloshing systems such as the Centaurus cluster, region\mbox{-}to\mbox{-}region bulk motions are observed \citep{xrismcollaboration2025}, and in PKS~0745$-$191, the appearance of paired cold fronts has been interpreted as suggestive of sloshing \citep{sanders2014}. 
AGN feedback may drive turbulence in the central core, but its influence remains unclear because the scale dependence of the turbulence velocity is not examined in this study.
Regarding sloshing, it is suggested that the velocity is less than $\rm 190\ km\ s^{-1}$ at each region.

We discuss the velocity structure of the ICM by comparing observations with RS simulations. In Section~\ref{sec:RS_Sim} we presented RS simulations under the assumption of a spatially constant turbulent velocity, $\sigma_{\rm turb}$, and zero bulk velocity. We assess velocity profile by jointly comparing the observed and simulated He$\alpha$ w/z line ratio, the line-broadening measurement of $\sigma_{\rm turb}$, and bulk velocity. We also examine how radial variations of $\sigma_{\rm turb}$ and the presence of bulk motions affect the observed w/z ratio and the measured line width. To this end, we compute a suite of models and confront them with the data.

The simulated w/z ratios and corresponding $\sigma_{\rm turb}$ are summarized in Figure~\ref{fig:wz_fourpanel}. 
The simulation results suggest that a low turbulent velocity is preferred in the central region, although higher turbulent velocities remain allowed.
In the outer region, the observed w/z ratio exhibits an enhancement relative to the optically thin case. However, our simulations show that the w/z ratio is largely insensitive to variations in $\sigma_{\rm turb}$ and $v_{\rm core}$ and remains consistent with the optically thin case.
These observational results show no contradiction with the emission line width and RS simulation results; further exposure time is required to discuss aspects such as the anisotropy of turbulent velocities.

Next, we explore the impact of bulk motions. As the $v_{\rm core}$ increases, RS becomes less effective and the w/z ratio approaches the optically thin limit (Figure~\ref{fig:wz_fourpanel}). 
The simulated w/z ratio prefers a low $v_{\rm core}$ in the central region, but the outer region is less sensitive to $v_{\rm core}$.

\subsection{Non-thermal pressure}
The non-thermal pressure fluctuation of the ICM is an important factor that affects the constraints on cosmological parameters. 
For the entire region, the temperature is $T = 5.866~\mathrm{keV}$, 
which corresponds to a sound speed of $c_{s} = (\gamma k_{\mathrm{B}} T / \mu m_{\mathrm{p}})^{1/2} = 1239~\mathrm{km~s^{-1}}$, 
where $\gamma = 5/3$ is the adiabatic index, $\mu = 0.61$ is the mean molecular. 
Assuming isotropic turbulence, the three-dimensional Mach number is 
$\mathcal{M}_{\mathrm{3D}} = \sqrt{3}\, v_{\rm turb} / c_{s} = 0.17$, for $v_{\rm turb} = 121^{+17}_{-17}~\mathrm{km~s^{-1}}$ 
within a field of view corresponding to $\approx350~\mathrm{kpc}$.
The ratio of non-thermal to total pressure, $P_{\mathrm{NT}} / P_{\mathrm{tot}}$, is given by \citep[e.g.,][]{eckert2019}
\begin{equation}
\frac{P_{\mathrm{NT}}}{P_{\mathrm{tot}}}
= \frac{\mathcal{M}_{\mathrm{3D}}^2}{\mathcal{M}_{\mathrm{3D}}^2 + 3 / \gamma}
= 1.5 \pm 0.3\%.
\end{equation}

The non-thermal pressure is similar to other galaxy clusters observed with XRISM and Hitomi (e.g., Perseus, Abell~2029, Coma, Ophiuchus).
XRISM observations of Abell~2029 measure a turbulent velocity of $169\pm10~\mathrm{km\ s^{-1}}$ and a non-thermal pressure fraction of $0.026\pm0.004$ within a FoV corresponding to $\approx 270~\mathrm{kpc}$ \citep{xrismcollaboration2025b}. Abell~2029 is similar to PKS~0745$-$191 in global properties, with $M_{500}=6.8\pm0.2$ (vs.\ $7.3\pm0.8$ for PKS~0745$-$191; \citealt{richard-laferriere2020}) and a core temperature of $\approx 7$~keV.
No clear central X-ray cavity has been detected in Abell~2029, and its central radio emission is weaker than that of PKS~0745$-$191, so comparing their non-thermal components is informative. \citet{richard-laferriere2020} report cavity powers of $8.7\times10^{43}~\mathrm{erg \ s^{-1}}$ for Abell~2029 and $1.7\times10^{45}~\mathrm{erg\ s^{-1}}$ for PKS~0745$-$191. Although there is a difference of about 20 times in the cavity power for each, the contribution of non-thermal pressure is only a few percent for both, so there is no significant difference.
Here, we have used the velocity dispersion across the entire FoV for simplicity, but in practice, comparisons must account for the scale of turbulent velocities.
The relationship between the central cavity, radio emission, and non-thermal pressure will likely be clarified by future XRISM observations.

\section{Conclusion}
\label{sec:conclusion}
We observed PKS~0745$-$191 with XRISM and determined ICM turbulent velocity, temperature and metallicity. 
The observed turbulent velocity in the central region is $78^{+34}_{-49} \rm {km\ s^{-1}}$ and the Fe \emissiontype{XXV} He$\alpha$ line is clearly divided into fine structure.
The Fe \emissiontype{XXV} He$\alpha$ w line in the central four pixels region ($\sim \rm 100\ kpc$) shows strong hints of suppression from optically thin model and the observed w/z ratio is $2.31^{+0.39}_{-0.35}$. PKS~0745$-$191 has a high electron density at the core, which could be caused by RS.
The redshift of the ICM is determined to be $z=0.10292\pm 0.00022$ and the upper limit of the bulk velocity is $190 \ \rm km\ s^{-1}$.
In order to estimate turbulent velocity and bulk velocity, we performed Monte-Carlo simulations of the RS effect.
The radial constant turbulent velocity case indicates preference the turbulent velocity is less than 250 $\rm km\ s^{-1}$, and the core bulk velocity $v_{\rm core}$ is less than 150 $\rm km\ s^{-1}$.
The obtained turbulent velocity from the RS simulation is consistent with the line broadening measurement.
This suggests the presence of the RS effect in PKS~0745$-$191. 
The results support the current understandings of the physical conditions of the ICM plasma obtained by the combinations of several X-ray observatories.

\section*{Funding}
The material is based upon work supported by NASA under award number 80GSFC21M0002.

\section*{Appendix 1. Reconstruction of the time dependent energy scale}
\label{sec:gain_tracking}
The gain of the Resolve calorimeter detectors is influenced by its environment including effects from the 50 mK heat sink temperature, bolometric loading from the cryogenic environment, and the temperature of the instrument electronics on the spacecraft panels. In normal operations, the gain is tracked using on-board $^{55}$Fe radioactive sources located on the instrument filter wheel, producing Mn K X-rays at 5.9 and 6.4~keV. These are rotated into the aperture periodically to provide gain fiducials and then the fiducials are linearly interpolated to reconstruct the instrument energy scale \citep{Porter2025}. The efficacy of this process is monitored using a calibration pixel that is part of the main detector array but located outside the instrument field of and continuously illuminated by a heavily collimated $^{55}$Fe radioactive source. The calibration pixel gain is reconstructed using the same fiducial intervals as the main array and this can then be compared to the result using a reconstruction from the continuous illumination. The average error using this process is about 0.17 eV over the first 18 months of observations \citep{Kelley2025}.

This observation was conducted early in the commissioning phase of the observatory, before the standard observation strategy was established. Thus, we must use an alternative strategy for reconstructing the instrument gain. During this observation, there were periodic gain fiducials using an on-board MXS producing Cr K and Cu K X-rays \citep{shipman2024} in a similar manner to the nominal procedure with the $^{55}$Fe sources on the instrument filter wheel. Two effects make this more complicated than just reconstructing the gain using a different fiducial line (Cr K$\alpha$ or Cu K$\alpha$ instead of Mn K$\alpha$). First, the MXS provides uneven illumination with the instrument gate valve closed since it partially obscures the source. The MXS is brightest along one edge of the array and then becomes progressively dimmer across the array. The final two rows of pixels (see Figure \ref{fig:image}) have insufficient counts for use in reconstructing the gain and have to be excluded from the analysis. Second, the intensity of the MXS was set too high for an accurate measurement of the line energy. The MXS is operated in short bursts with a relatively low duty cycle and this can result in high instantaneous count rates and can adversely affect the measured line position during the short MXS exposures \citep{sawada2025}. Luckily there was a long time period at the end of the observation, 12 hours, where both the MXS and the FW were operated simultaneously, allowing for the calibration of the line offsets to be back propagated into the observation of the celestial source. 

\begin{figure}[htbp]
\includegraphics[width=0.48\textwidth]{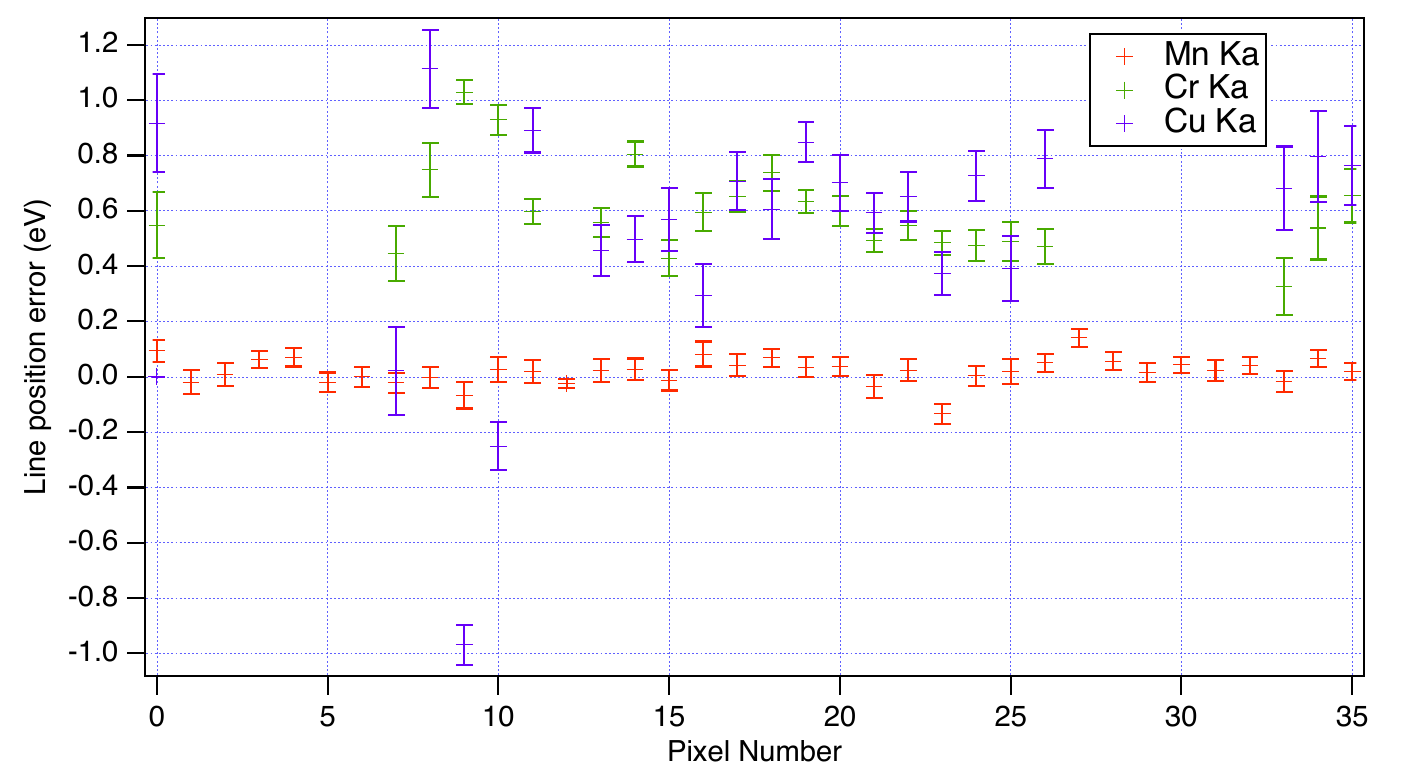}
\caption{The line position errors of the MXS Cr K$\alpha$, and Cu K$\alpha$ lines after gain reconstruction using the FW $^{55}$Fe source (Mn K$\alpha$). The MXS lines are displaced due to operating the source at relatively high instantaneous intensity. The Mn K$\alpha$ lines are unaffected. Missing values are for pixels where the MXS flux was insufficient to provide a measurement.
{Alt text: Single line graph showing the line position errors of each fiducial line. The x axis shows pixel ID of Resolve. The y axis shows the line position error in eV.}
}
\label{fig:app_figureA}
\end{figure}

\begin{figure}[htbp]
\includegraphics[width=0.48\textwidth]{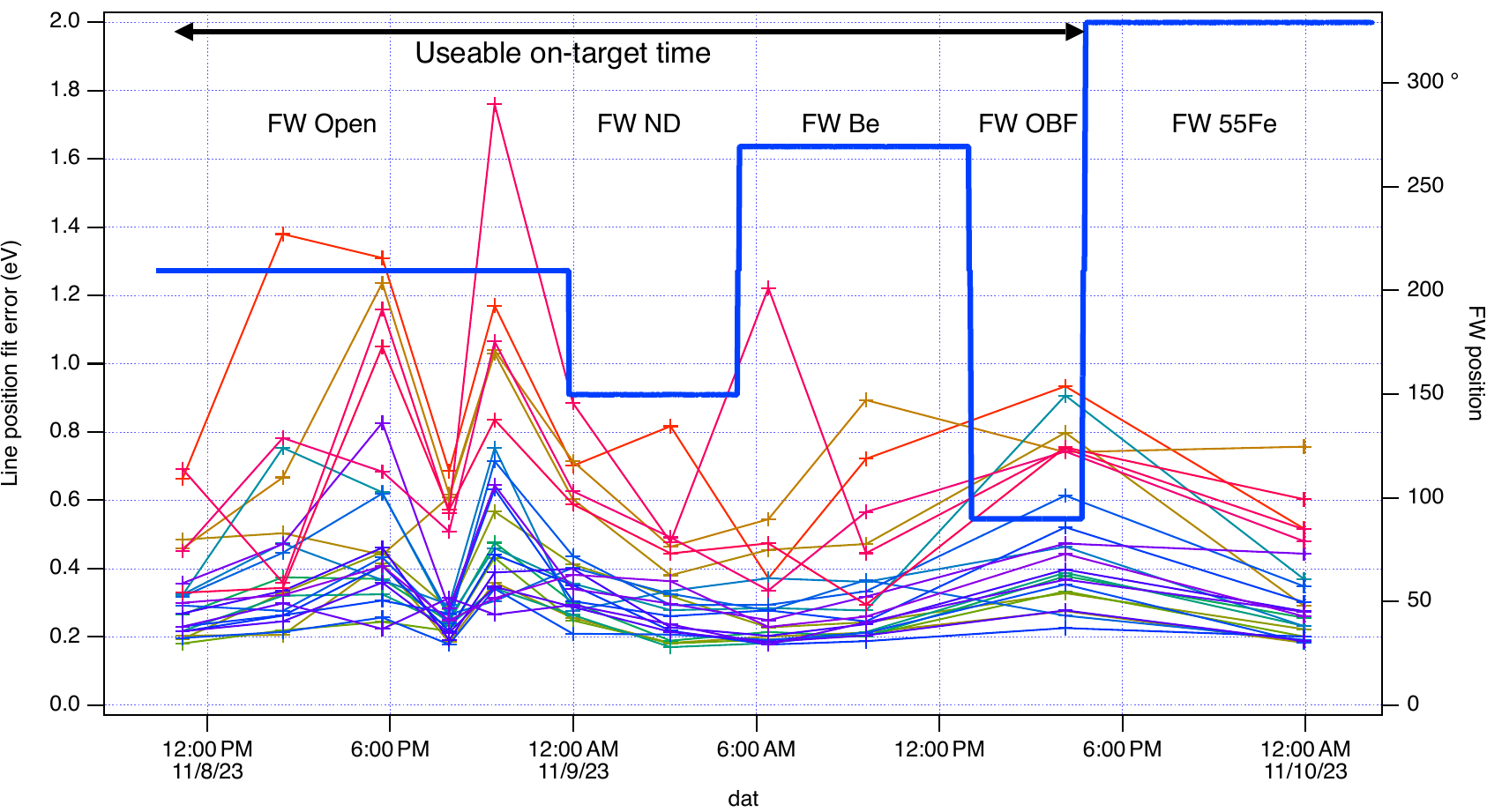}
\caption{The statistical uncertainty in the fits to the Cr Ka gain fiducials during the observation. Each pixel is a different color and the lines are to guide the eye. There are 11 fiducial intervals, and the differing uncertainties reflect the uneven MXS illumination across the detector array. The different FW positions are also shown by the heavy blue line and the right axis.
{Alt text: Line graph showing the statistical uncertainty of gain fiducials for each Resolve pixel during the observation.}
}
\label{fig:app_figureB}
\end{figure}

\begin{figure}
\centering
\includegraphics[width=0.48\textwidth]{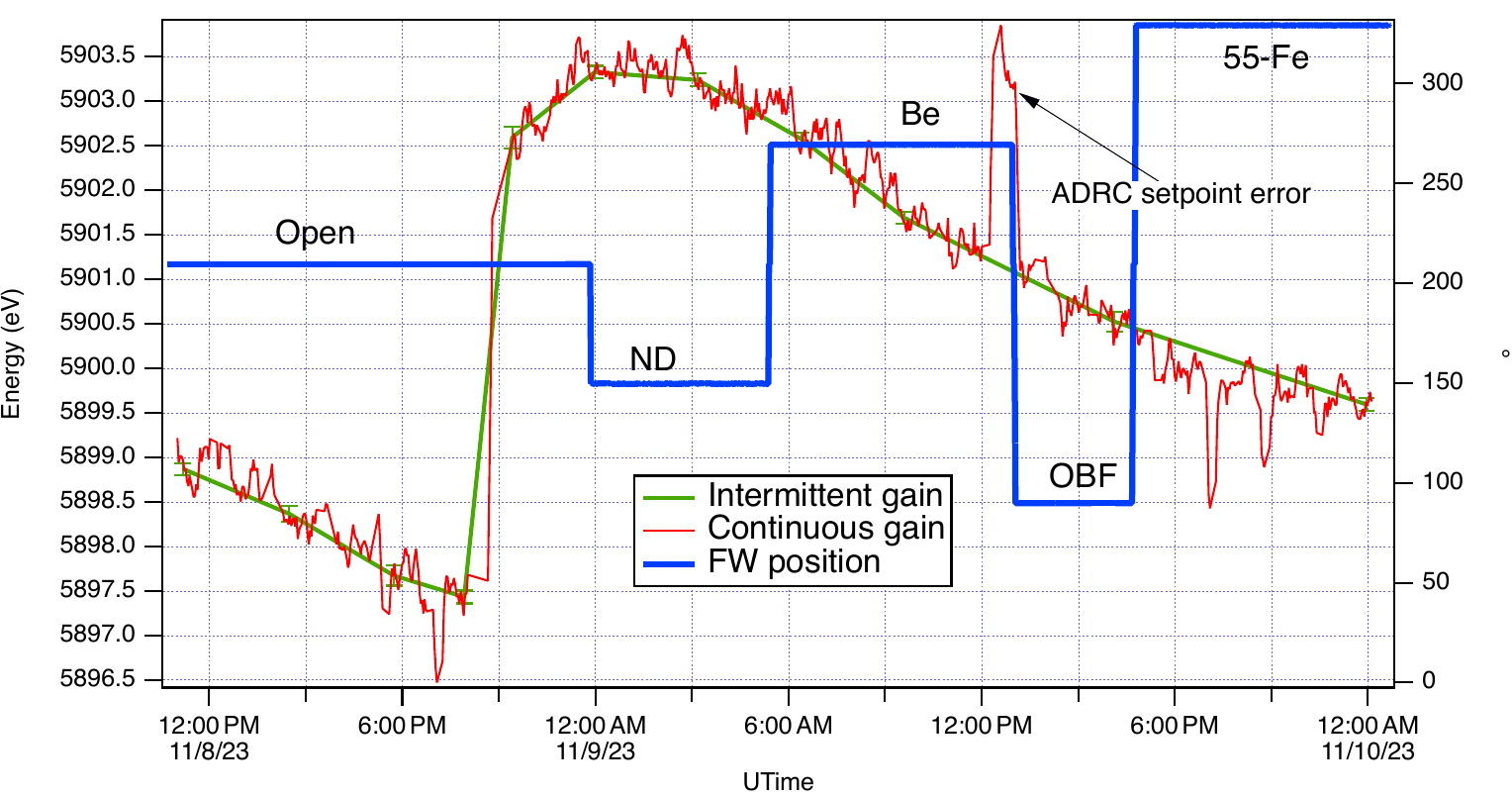}
\caption{The calibration pixel gain as measured by the same gain fiducial intervals as the main observation (green), and a continuous measurement (red). This shows the efficacy of the standard gain reconstruction process (green) compared to a representation of the true underlying function (red). Variations in the red curve are predominately from orbital variations in the temperature of the room temperature electronics. The large excursion around 10p.m. on 2023/11/23 is from a recycle of the instrument 50 mK cooler and the excursion marked “ADRC setpoint error” is due to a rare error in the 50 mK temperature controller that occurred early in the mission. This interval was excluded during the analysis presented here.
{Alt text: Single line graph showing the gain fiducials measured by the same gain fiducial intervals as the main observation, and a continuous measurement.}
}
\label{fig:app_figureC}
\end{figure}

The line position errors for the Cr K$\alpha$ and Cu K$\alpha$ lines after gain reconstruction using Mn K$\alpha$ from the FW $^{55}$Fe sources are shown in Figure \ref{fig:app_figureA}. We used the brighter Cr K$\alpha$ line for the gain fiducials during the observation of PKS~0745$-$191. The statistical uncertainties for the gain fiducial measurements during the observation are shown in Figure \ref{fig:app_figureB} and vary strongly across the array due to the uneven illumination from the MXS. Simultaneously, we measured the calibration pixel during these same fiducial intervals and measured the gain reconstruction error against the calibration pixel excluding the gain fiducial intervals (see figure \ref{fig:app_figureC}). The resultant gain history was then inserted into the standard Resolve pipeline processing and the standard energy scale reconstruction process was utilized. 

The uncertainty in the absolute energy scale was estimated for each position of the FW during this observation in the following manner. The standard Resolve energy scale uncertainty is estimated by summing the cal pixel error during the observation which monitors the efficacy of the sparse sampling of the time dependent gain function, with an estimate of the systematic uncertainty in the full energy scale, likely dominated by our knowledge of the underlying line shapes of the fiducial lines. In the region 5.4--8.0 keV this is estimated at 0.3 eV \citep{eckart2024}. For this observation, we have several additional sources of uncertainty: the statistical uncertainty in the gain fiducial measurements, and the measurement of the energy scale offsets in the MXS (Figure \ref{fig:app_figureB}). The systematic error is then the uncorrelated sum of the statistical uncertainty in the Cr K$\alpha$ offsets ($\sim$0.2 eV), the statistical uncertainty in the Cr K$\alpha$ fiducials, the error in the cal pixel reconstruction, and the systematic uncertainty in the underlying energy scale (0.3 eV). The end result is that the 5.4--8.0 keV uncertainty for each FW interval is (0.62, 0.62, 0.58, 0.57, and 0.58 eV) for the intervals (FW open, ND, Be, OBF, and entire observation) respectively.

\section*{Appendix 2. Model of the NXB}
\label{sec:nxb_model}
The NXB spectrum was modeled using a phenomenological model\footnote{see \url{https://heasarc.gsfc.nasa.gov/docs/xrism/analysis/nxb/nxb_spectral_models.html}}.
The normalization of the continuum component and those of the individual Gaussian lines were fixed to their best-fit values.
The emission lines included in the model were Al-K$\alpha_{1,2}$, Au-M$\alpha_{1}$, Cr-K$\alpha_{1,2}$, Mn-K$\alpha_{1,2}$, Fe-K$\alpha_{1,2}$, Ni-K$\alpha_{1,2}$, Cu-K$\alpha_{1,2}$, Au-L$\alpha_{1,2}$, and Au-L$\beta_{1,2}$.
The $\alpha_{1}$-to-$\alpha_{2}$ intensity ratios were fixed at 2:1.
The best-fit parameters of the NXB model for the entire region in the non-SSM case are presented in Table~\ref{tab:nxb_gauss_norm}.

\begin{table}[htbp]
\centering
\caption{Best-fit parameters of the phenomenological NXB model. We showed the overall scaling factor and normalization of the \texttt{gaussian} components. The normalization of the {\tt gaussian} component represents 
the total line flux in units of photons cm$^{-2}$ s$^{-1}$.}
\label{tab:nxb_gauss_norm}
\begin{threeparttable}
\begin{tabular}{l c c}
\toprule
Component & Line Energy & Normalization \\
\midrule
Overall scaling factor & --- & 0.9325 \\

Al--K$\alpha_1$ & 1.4867 & $2.58\times10^{-5}$ \\
Au--M$\alpha_1$ & 2.1229 & $4.58\times10^{-5}$ \\
Cr--K$\alpha_1$ & 5.4147 & $1.50\times10^{-5}$ \\
Mn--K$\alpha_1$ & 5.8980 & $4.26\times10^{-5}$ \\
Fe--K$\alpha_1$ & 6.4038 & $8.17\times10^{-6}$ \\
Ni--K$\alpha_1$ & 7.4782 & $3.83\times10^{-5}$ \\
Cu--K$\alpha_1$ & 8.0478 & $1.86\times10^{-5}$ \\
Au--L$\alpha_1$ & 9.6280 & $1.98\times10^{-5}$ \\
Au--L$\alpha_2$ & 9.7133 & $1.54\times10^{-4}$ \\
Au--L$\beta_1$  & 11.4423 & $1.28\times10^{-4}$ \\
Au--L$\beta_2$  & 11.5847 & $4.85\times10^{-5}$ \\

\bottomrule
\end{tabular}
\end{threeparttable}
\end{table}

\section*{Appendix 3. Differences between AtomDB 3.0.9 and 3.1.3}
In the spectrum analysis of Section \ref{sec:spec}, we adopted \textsc{AtomDB} v3.0.9 to produce the CIE model.
While writing the paper, a few updates were made to AtomDB, whose latest version is 3.1.3 at the time of writing. primarily correcting the emissivity of dielectric-recombination satellite lines.
This may affect the measurement results of the w/z ratio, therefore we examined \textsc{AtomDB} v3.1.3 by adopting it in the spectral fit in the non-SSM analysis.
The fitting results are shown in the table~3. Parameters including the w/z ratio show slight variations within the statistical error range, but these differences are negligible in our analysis.
\begin{table*}[htbp]
  \begin{threeparttable}
  \caption{Best-fit parameters of each region spectrum (Non-SSM only, single-temperature fit, \textsc{AtomDB} v3.1.3). Quoted errors are $1\sigma$.}
  \label{tab:best-fit-single-3.1.3}
  \centering
  \begin{tabular*}{\textwidth}{@{\extracolsep{\fill}} l lll}
    \hline
    & \multicolumn{3}{c}{\textbf{Non-SSM}} \\[-1pt]
    \cline{2-4}
    Parameter & Entire & Central & Outer \\ \hline\hline
    \multicolumn{4}{l}{\texttt{bvapec (exclude w,z) + zgauss(w) + zgauss(z), AtomDB 3.1.3}} \\[2pt]
    $kT$ (keV)                & $5.827^{+0.094}_{-0.092}$ & $5.29^{+0.13}_{-0.12}$ & $6.19^{+0.13}_{-0.14}$ \\
    $z_{\mathrm{Open}}$      & $0.102946^{+4.4\times10^{-5}}_{-6.0\times10^{-5}}$ & $0.10285^{+5.0\times10^{-5}}_{-1.1\times10^{-4}}$ & $0.102997^{+5.8\times10^{-5}}_{-8.6\times10^{-5}}$ \\
    $z_{\mathrm{OBF}}$       & $0.10291^{+8.7\times10^{-5}}_{-9.8\times10^{-5}}$ & $0.10288^{+1.1\times10^{-4}}_{-1.3\times10^{-4}}$ & $0.10290^{+1.3\times10^{-4}}_{-1.3\times10^{-4}}$ \\
    $z_{\mathrm{ND}}$        & $0.10304^{+1.5\times10^{-4}}_{-1.4\times10^{-4}}$ & $0.10294^{+2.8\times10^{-4}}_{-1.8\times10^{-4}}$ & $0.10301^{+1.9\times10^{-4}}_{-2.7\times10^{-4}}$ \\
    $z_{\mathrm{Be}}$        & $0.102968^{+8.6\times10^{-5}}_{-4.2\times10^{-5}}$ & $0.103039^{+9.2\times10^{-5}}_{-9.3\times10^{-5}}$ & $0.10290^{+1.1\times10^{-4}}_{-7.0\times10^{-5}}$ \\
    $\sigma_{\mathrm{turb}}$ (km\,s$^{-1}$) & $128^{+17}_{-17}$ & $129^{+28}_{-29}$ & $131^{+24}_{-22}$ \\
    $Z_{\mathrm{Fe}}$ ($Z_{\odot}$) & $0.421^{+0.022}_{-0.021}$ & $0.442^{+0.035}_{-0.034}$ & $0.403^{+0.028}_{-0.026}$ \\
    $Z_{\mathrm{Ni}}$ ($Z_{\odot}$) & $0.39^{+0.11}_{-0.10}$ & $0.38^{+0.17}_{-0.15}$ & $0.38^{+0.14}_{-0.13}$ \\
    apec norm$^{*}$           & $0.0675^{+8.1\times10^{-4}}_{-8.1\times10^{-4}}$ & $0.09607^{+2.3\times10^{-4}}_{-3.3\times10^{-4}}$ & $0.05747^{+7.6\times10^{-4}}_{-6.9\times10^{-4}}$ \\
    $\mathrm{norm}_w^{\dagger}$ ($10^{-5}$) & $7.53^{+0.44}_{-0.42}$ & $10.63^{+0.96}_{-0.92}$ & $6.44^{+0.48}_{-0.47}$ \\
    $\sigma_w$ (eV)          & $4.71^{+0.33}_{-0.31}$ & $3.94^{+0.45}_{-0.46}$ & $5.13^{+0.54}_{-0.40}$ \\
    $\mathrm{norm}_z^{\dagger}$ ($10^{-5}$) & $2.65^{+0.28}_{-0.27}$ & $4.31^{+0.64}_{-0.61}$ & $2.03^{+0.30}_{-0.27}$ \\
    $\sigma_z$ (eV)          & $3.34^{+0.55}_{-0.49}$ & $2.53^{+0.69}_{-0.66}$ & $3.42^{+0.84}_{-0.85}$ \\
    w/z ratio               & $2.84^{+0.33}_{-0.34}$  & $2.47^{+0.42}_{-0.42}$   & $3.18^{+0.49}_{-0.53}$  \\
    optically thin            & 3.24 & 3.16  &  3.30 \\
    cstat/d.o.f               & 5703.24/6220 & 3239.59/4018 & 4496.75/5166 \\
    \hline
  \end{tabular*}
  \begin{tablenotes}
  \item[*] Normalization of the \texttt{bvapec} model. Defined by $\frac{10^{-14}}{4\pi [D_A(1+z)]^2 }\int n_e n_H dV$, where $D_A$ is the angular diameter distance.
  \item[$\dagger$] Integrated flux of the \texttt{zgauss} model.
  \end{tablenotes}
  \end{threeparttable}
\end{table*}

\begin{table*}[htbp]
  \begin{threeparttable}
  \caption{Best-fit parameters of each region spectrum (Non-SSM only, multi-temperature fit). Quoted errors are $1\sigma$.}
  \label{tab:best-fit-double-3.1.3}
  \centering
  \begin{tabular*}{\textwidth}{@{\extracolsep{\fill}} l lll}
    \hline
    & \multicolumn{3}{c}{\textbf{Non-SSM}} \\[-1pt]
    \cline{2-4}
    Parameter & Entire & Central & Outer \\ \hline\hline
    \multicolumn{4}{l}{\texttt{bvapec (cool) + bvapec (hot), AtomDB 3.0.9}} \\[2pt]
    $kT_{\rm cool}$ (keV)  & $3.83^{+0.27}_{-0.36}$   & $0.44^{+0.18}_{-0.23}$   & $4.22^{+0.27}_{-0.34}$ \\
    $kT_{\rm hot}$ (keV)   & $8.55^{+0.73}_{-0.55}$   & $5.48^{+0.14}_{-0.14}$   & $8.62^{+0.30}_{-0.28}$ \\
    apec norm$_{\rm cool}^{*}$      & $0.0347^{+0.0061}_{-0.0063}$ & $0.300^{+3.700}_{-0.200}$ & $0.0289^{+0.0024}_{-0.0025}$ \\
    apec norm$_{\rm hot}^{*}$       & $0.0364^{+0.0059}_{-0.0064}$ & $0.0937^{+0.0013}_{-0.0014}$ & $0.0307^{+0.0018}_{-0.0016}$ \\
    $z_{\rm Open}$         & $0.102919^{+5.5\times10^{-5}}_{-5.1\times10^{-5}}$ & $0.102859^{+5.2\times10^{-5}}_{-9.9\times10^{-5}}$ & $0.102971^{+7.4\times10^{-5}}_{-7.2\times10^{-5}}$ \\
    $z_{\rm OBF}$          & $0.102960^{+6.0\times10^{-5}}_{-1.4\times10^{-4}}$ & $0.102870^{+1.5\times10^{-4}}_{-1.0\times10^{-4}}$ & $0.102890^{+1.0\times10^{-4}}_{-1.7\times10^{-4}}$ \\
    $z_{\rm ND}$           & $0.103050^{+1.3\times10^{-4}}_{-1.6\times10^{-4}}$ & $0.103010^{+2.2\times10^{-4}}_{-2.4\times10^{-4}}$ & $0.103010^{+1.8\times10^{-4}}_{-2.6\times10^{-4}}$ \\
    $z_{\rm Be}$           & $0.102999^{+7.2\times10^{-5}}_{-5.2\times10^{-5}}$ & $0.103072^{+8.1\times10^{-5}}_{-9.8\times10^{-5}}$ & $0.102944^{+8.5\times10^{-5}}_{-8.7\times10^{-5}}$ \\
    $\sigma_{\rm turb}$ ($\rm km\ s^{-1}$) & $149.0^{+11.0}_{-11.0}$ & $127.0^{+15.0}_{-18.0}$ & $158.0^{+16.0}_{-15.0}$ \\
    $Z_{\mathrm{Fe}}$ ($Z_{\odot}$) & $0.436^{+0.016}_{-0.015}$ & $0.414^{+0.023}_{-0.022}$ & $0.423^{+0.020}_{-0.019}$ \\
    $Z_{\mathrm{Ni}}$ ($Z_{\odot}$) & $0.430^{+0.120}_{-0.110}$ & $0.390^{+0.170}_{-0.150}$ & $0.410^{+0.160}_{-0.150}$ \\
    cstat/dof         & 5709.56/6222 & 3241.86/4020 & 4500.59/5168 \\
    \hline
  \end{tabular*}
  \begin{tablenotes}
  \item[*] Normalization of the \texttt{bvapec} model. Defined by $\frac{10^{-14}}{4\pi [D_A(1+z)]^2 }\int n_e n_H dV$, where $D_A$ is the angular diameter distance.
  \end{tablenotes}
  \end{threeparttable}
\end{table*}

\section*{Appendix 4. Multi-temperature model}
\label{sec:multitemp}
We fitted the spectra with \texttt{bvapec}\textsubscript{hot}+\texttt{bvapec}\textsubscript{cool}
 components in the non-SSM case.
The turbulent velocity, redshift, and metal abundance of the cool component are linked to the hot component. The fitting result is shown in Table~4. The changes of Cash statistics in comparison with those Table 1 are less than 10 in > 4000 bins. The best-fit temperature of entire region and outer region are close to the SSM case, and we can understand that we see the mutual fluxes of each region. In the central region, the cool component temperature is $0.44^{+0.18}_{-0.23}$ keV.
At $0.44\,\mathrm{keV}$, Fe~\emissiontype{XVII} dominates the ion fraction ($\sim70\%$), and the ratio Fe~$\emissiontype{XXV}$/Fe~$\emissiontype{XVII}$ is below $10^{-9}$, implying that emission from cool component is negligible at this temperature.
No higher temperature component radiating Fe \emissiontype{XXV} He$\alpha$ w was detected, suggesting that the decrease in w is due to RS.

\section*{Appendix 5. Non-SSM case simulation}
\label{sec:nonssm_sim}
For the robust analysis, we performed RS simulation in non-SSM case.
Because we observed mutual fluxes of each region, we need to take into accont the SSM effects in the RS simulation.
We adapted \texttt{Heasim} tools which can simulate the Resolve observation by taking into account the PSF and detector response for our RS simulation spectra. 
We used the same RMF and ARF files as the observation data, and the PSF from the file \texttt{sxs\_psfimage\_20140618.fits} distributed as part of the simulation tools\footnote{See, e.g., \url{https://heasarc.gsfc.nasa.gov/FTP/xrism/prelaunch/simulation/sim3/}}.

\begin{figure*}[htbp]
  \centering
  \begin{minipage}[b]{0.45\textwidth}
    \centering
    \includegraphics[width=\linewidth]{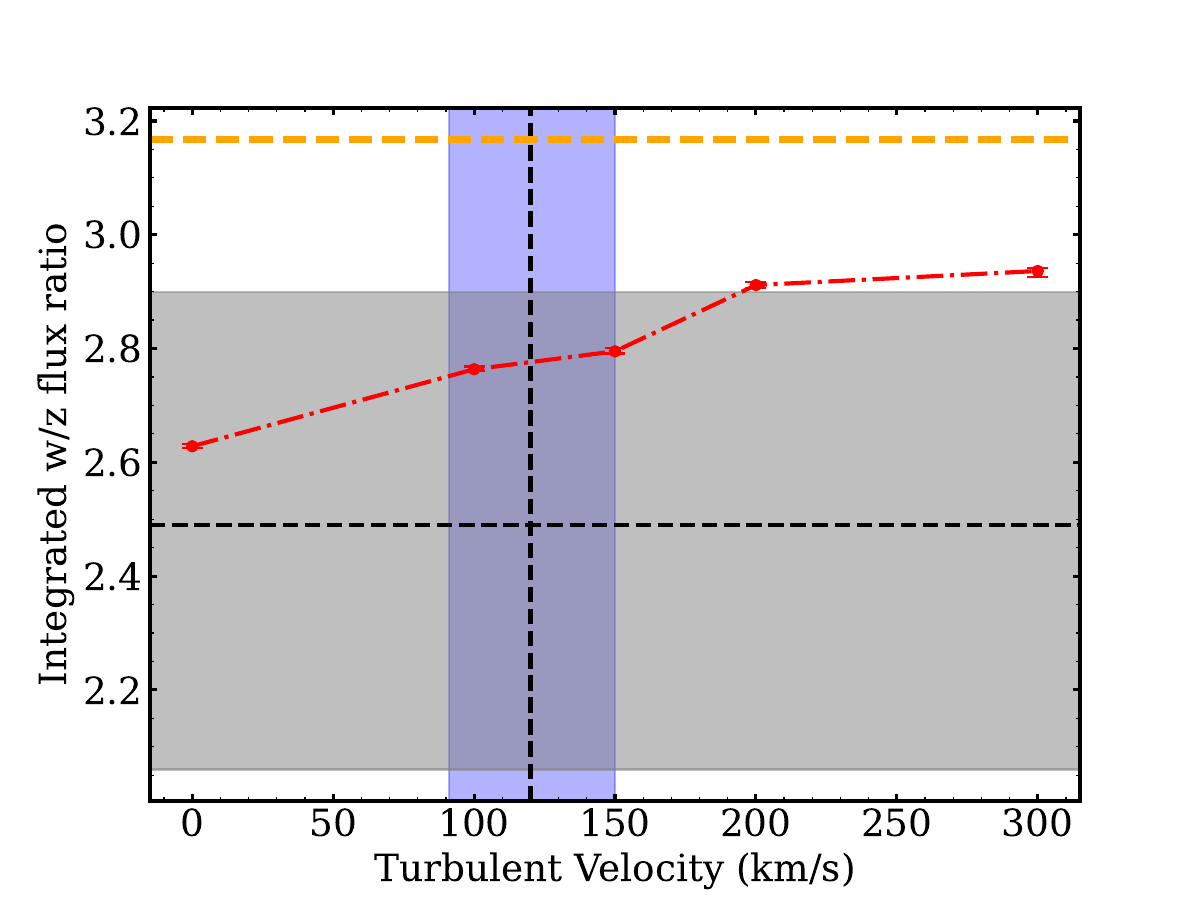}
  \end{minipage}
  \begin{minipage}[b]{0.45\textwidth}
    \centering
    \includegraphics[width=\linewidth]{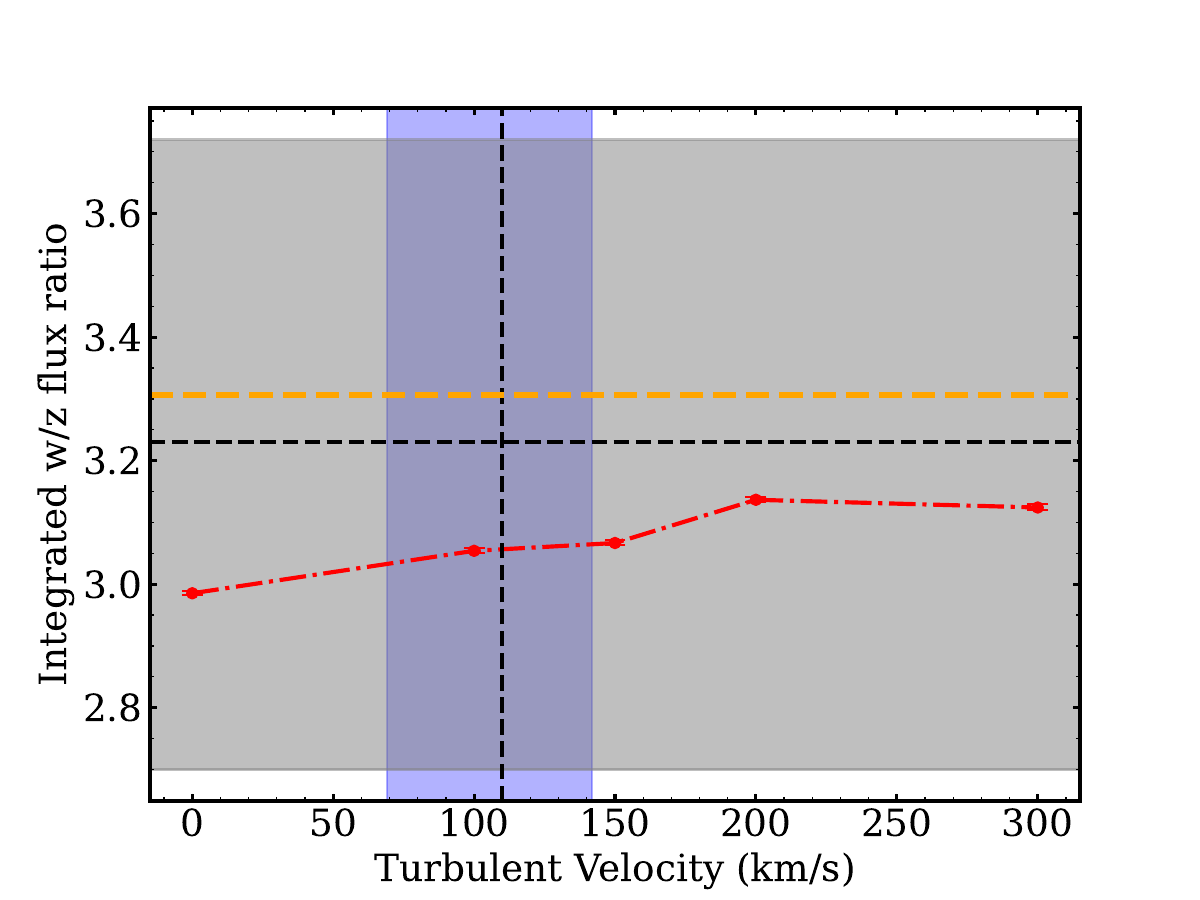}
  \end{minipage}

  \caption{Integrated w/z flux ratio as a function of turbulent velocity for each region. The red points correspond to $v_{bulk}=0 \ \rm km\ s^{-1}$. The black dashed lines are the measured w/z flux ratio and turbulent velocity obtained by \texttt{bvapec + zgauss(w) + zgauss(z)} model in non-SSM case. The gray and blue spans are the 1$\sigma$ uncertainty. Orange dashed lines are the w/z flux ratio of the optically thin case.
  (left): Simulation result in the central region; (right): Simulation result in the outer region.   
  {Alt text: Two line graphs showing the comparison of w/z flux ratio as a function of turbulent velocity. The x axis shows turbulent velocity.}
  }
  \label{fig:wz_fourpanel_nonssm}
\end{figure*}

We showed the simulation result with the constant turbulent velocity and bulk velocity is zero case in Figure \ref{fig:wz_fourpanel_nonssm}.
The estimated turbulent velocity is less than 200 $\rm km\ s^{-1}$, and it is consistent with the SSM case.

\bibliography{../cls_bib/PASJ_PKS}

\end{document}